\documentclass[aip,
jcp,
reprint,
superscriptaddress,
amsmath,amssymb,floatfix,
longbibliography
]{revtex4-2}
\usepackage{graphicx}
\usepackage{dcolumn}
\usepackage{comment}
\usepackage{bm}
\usepackage{xcolor}
\definecolor{darkblue}{rgb}{0,0,0.6}
\usepackage[normalem]{ulem}
\usepackage{hyperref}
\hypersetup{colorlinks=true, linkcolor=darkblue, citecolor=darkblue, urlcolor=darkblue}

\definecolor{lightgray}{gray}{0.8}

\begin{document}

\title{Kinetic Theory of Chiral Active Disks: Odd Transport and Torque Density}

\author{Rapha\"el Maire}
\thanks{\href{mailto:maire@ub.edu}{maire@ub.edu}}
\affiliation{Department of Condensed Matter, University of Barcelona, 08028 Barcelona, Spain}

\author{Alessandro Petrini}
\affiliation{Sapienza University of Rome, IT-00185 Rome, Italy}

\author{Umberto Marini Bettolo Marconi}
\affiliation{Universit\`a di Camerino, Scuola di Scienze e Tecnologie, 62032, Camerino, Italy}

\author{Lorenzo Caprini}
\thanks{\href{mailto:lorenzo.caprini@uniroma1.it}{lorenzo.caprini@uniroma1.it}}
\affiliation{Sapienza University of Rome, IT-00185 Rome, Italy}

\date{\today}

\begin{abstract}
    Parity-odd transport is a central signature of chiral fluids, yet analytical predictions are sparse. Here, we introduce a minimal two-dimensional hard-disk gas in which chirality arises solely from a collision-induced transverse impulse. Motivated by granular spinners, collisions are dissipative and inject orbital angular momentum through a fixed tangential ``kick'' at contact. Starting from a Boltzmann-Enskog description, we derive nonlinear hydrodynamic equations for density, momentum, and temperature, and show that chirality generates an antisymmetric homogeneous stress corresponding to a nonzero torque density. In the dilute limit, a Chapman-Enskog expansion yields analytical predictions for transport coefficients, including odd viscosity, odd thermal conductivity, and odd self-diffusivity, in good agreement with numerical simulations. This minimal kinetic model can serve as a foundation for systematic coarse-graining of chiral fluids and as a tractable benchmark for gaining insight into odd transport across a broader class of chiral systems.
\end{abstract}

\maketitle 

\section{Introduction}

Chiral active matter encompasses a broad class of systems that break parity symmetry at the level of their microscopic constituents~\cite{lowen2016chirality,liebchen2022chiral}. Representative examples range from microswimmers---such as sperm cells and bacteria moving along circular trajectories near surfaces~\cite{woolley2003motility,petroff2015fast}---to microorganisms, including cells~\cite{xu2007polarity}, algae~\cite{huang2021circular}, and starfish embryos~\cite{tan2022odd}, which exhibit the formation of rotating clusters. Chiral particles have also been artificially engineered by designing micro- or macroscopic objects with broken rotational symmetry. Notable examples include chiral active colloids moving along circular paths~\cite{kummel2013circular} and granular spinners that self-rotate due to external driving mechanisms such as light~\cite{siebers2023exploiting}, internal motors~\cite{carrillo2025depinning}, vibrating plates~\cite{scholz2018rotating,caprini2025spontaneous,tiwari2026reentrant,huang2023odd}, or airflows~\cite{lopez2022chirality,vega2022diffusive}.

A wealth of distinctive phenomena characterizes these chiral systems \cite{caprini2019active,levis2019activity,kreienkamp2022clustering,bickmann2022analytical,pisegna2025spinning}. In particular, a circularly moving object that breaks parity symmetry typically exhibits edge currents along confining boundaries~\cite{workamp2018symmetry,mecke2024emergent,negi2023geometry,caprini2025active}, circulating currents in the presence of an external confining potential~\cite{caprini2023chiral}, as well as rotating crystallites in solid structures~\cite{huang2020dynamical,musacchio2026circling}. These effects originate from their non-standard diffusive properties, which couple different Cartesian components of the motion. As a consequence, Fick's law governing the density evolution involves a diffusion matrix with antisymmetric off-diagonal components. These terms are commonly referred to as odd diffusion~\cite{hargus2021odd,caprini2025active,vega2022diffusive, abdoli2026dynamicaldensityfunctionaltheory, faedi2026mobilitybasedapproachtransport}, or odd mobilities~\cite{reichhardt2019active, reichhardt2022active} and have been analytically predicted starting from microscopic models of chiral active particles~\cite{hargus2021odd,kalz2022collisions,kalz2024oscillatory}. Related diffusive responses and mobilities have also been reported in magnetic skyrmions due to gyrotropic effects~\cite{everschor2014real,Weibenhofer_Rozsa_Nowak_2021,dohi2023enhanced}.

More broadly, the behavior of chiral systems is often described using hydrodynamic theories for the density and momentum fields, reminiscent of generalized Navier–Stokes equations operating far from equilibrium. In contrast to conventional fluids, the viscosity tensor of a two-dimensional chiral system contains antisymmetric components, known as odd viscosity coefficients (see Ref.~\onlinecite{fruchart2023odd} for a review). Odd viscosity~\cite{fruchart2023odd, banerjee2017odd, markovich2021odd, markovich2025chiral} was first introduced in hydrodynamic theory by Avron~\cite{avron1998odd} and has recently been observed experimentally both in active-matter systems~\cite{soni2019odd} and in electron fluids~\cite{berdyugin2019measuring}. Unlike bulk and shear viscosity, odd viscosity does not dissipate energy, yet it can induce Hall-like transport~\cite{lou2022odd} and lift forces~\cite{lier2023lift, PhysRevLett.131.178303}.

Although the recent literature on the hydrodynamics of chiral fluids is rapidly growing \cite{fruchart2023odd, huang2025anomalous}, derivations or analytical predictions of odd transport coefficients remain scarce. Recently, hydrodynamic equations for chiral active fluids were derived~\cite{Marconi2026hydrodynamics} from the Langevin dynamics of chiral particles subject to odd interactions~\cite{caprini2025Bubble}, i.e., effective transverse forces that generate a net torque~\cite{caporusso2024phase,caprini2025modeling,digregorio2025phase,guo2026tuning}. This work showed that chiral fluids are not only characterized by odd viscosity---as assumed in most hydrodynamic treatments---but also by a chirality-induced torque density~\cite{lee2025odd, braverman2021topological, fruchart2023odd}. However, odd viscosity is often introduced phenomenologically, on symmetry grounds, rather than derived microscopically from the underlying many-body physics. Notable exceptions are Ref.~\onlinecite{eren2025collisional}, which used linear-response theory to obtain an interaction-induced odd viscosity in the dilute limit for rotating particles undergoing binary collisions, and Ref.~\onlinecite{lier2026chapman}, which developed a Chapman--Enskog theory of energy- and momentum-conserving chirally colliding hard disks.

Here, we present one of the first analytical studies predicting transport coefficients for chiral fluids. We derive explicit expressions for the torque density, odd viscosity, odd thermal conductivity, and odd self-diffusivity within a kinetic theory in which chirality arises solely from the collision-induced injection of orbital angular momentum. Motivated by granular systems, particle collisions are dissipative yet inject energy transversely to the line connecting the particle centers (see Fig.~\ref{fig: collision} for an illustration). This collision rule is the instantaneous analog of odd interactions, where a transverse interparticle force represents, at a coarse-grained level, the effective coupling between two rotating particles. Such forces may be mediated by hydrodynamic interactions in a surrounding fluid or arise from the rotational friction of granular spinners.

We introduce the model in Sec.~\ref{sec: model}, summarize the main results in Sec.~\ref{sec: results overview}, and present the derivations in Sec.~\ref{sec: results derivation}. We conclude with a discussion.

\section{Model}\label{sec: model}

We consider a chiral active fluid consisting of $N$ hard disks of diameter $\sigma$ and mass $m$, moving in a box of size $L$ with periodic boundary conditions.
Motivated by kinetic models typical of driven granular gases, particles move ballistically until they undergo binary collisions, where chirality manifests through an injection of angular momentum.
Specifically, when two particles $1$ and $2$ come into contact ($|\mathbf r_1-\mathbf r_2|=\sigma$), their velocities are updated via an active, parity-breaking collision rule:
\begin{subequations}
\begin{flalign}
    \bm v_1'&= \bm v_1 - \dfrac{1+\alpha}{2}(\bm v_{12}\cdot \hat{\bm\sigma}_{12})\hat{\bm\sigma}_{12} -  \Delta\hat{\bm\sigma}_{12}^\perp\,, \\
    \bm v_2'&= \bm v_2 + \dfrac{1+\alpha}{2}(\bm v_{12}\cdot \hat{\bm\sigma}_{12})\hat{\bm\sigma}_{12} + \Delta\hat{\bm\sigma}_{12}^\perp\,,
\end{flalign}
\label{eq: collision rule}
\end{subequations}
where $\hat{\bm \sigma}_{12}=(\mathbf r_1-\mathbf r_2)/\sigma$,  $\bm v_{12}=\bm v_1-\bm v_2$ and $\bm v_{1}$ and $\bm v'_1$ are the velocities of particle $1$ before and after the collision, respectively.

The term $(1+\alpha)(\bm v_{12}\cdot\hat{\bm\sigma}_{12})\hat{\bm\sigma}_{12}/2$ corresponds to the inelastic hard-disk normal impulse. Like a repulsive force, this term reverses the normal component of the relative velocity and reduces its magnitude. Indeed, the post-collisional velocity is related to the pre-collisional one through the following relation involving the restitution coefficient $0\le\alpha\le1$:
\begin{equation}
(\bm v_{12}'\cdot\hat{\bm\sigma}_{12})=-\alpha(\bm v_{12}\cdot\hat{\bm\sigma}_{12})\,.
\end{equation}
Consequently, $\alpha=1$ corresponds to elastic hard-disk collisions without energy dissipation, while $\alpha<1$ dissipates kinetic energy at each collision.

The second term, $ \Delta \hat{\bm\sigma}_{12}^{\perp}$, is an active tangential impulse that breaks parity symmetry and is generated by a chiral mechanism. Here
$\hat{\bm\sigma}_{12}^\perp=\bm\varepsilon\cdot\hat{\bm\sigma}_{12}=(\hat \sigma_{12, y}, -\hat \sigma_{12, x})$ with $\bm\varepsilon$ the two-dimensional Levi-Civita symbol ($\varepsilon_{xy}=-\varepsilon_{yx}=1$). Thus $\Delta$ has the dimension of a velocity and sets the magnitude of a fixed transverse ``kick'' at contact, with its handedness controlled by $\mathrm{sign}(\Delta)$. Equivalently, this term corresponds to a singular nonconservative contact force with an odd (transverse) component, and it breaks the conservation of \emph{orbital} angular momentum in each collision, providing a microscopic source of chiral stresses. The equilibrium hard-disk limit corresponds to $\Delta\to 0$ and $\alpha = 1$, while $\Delta \to 0$ and $\alpha<1$ yield a freely cooling granular gas~\cite{brilliantov2010kinetic}. 

The collision rule can alternatively be expressed in terms of an impulsive force $\mathbf F_{12}$ acting on particle 1 due to particle 2:
\begin{equation}
    \mathbf F_{12}=-m\left(\dfrac{1+\alpha}{2}(\bm v_{12}\cdot \hat{\bm\sigma}_{12})\hat{\bm\sigma}_{12} + \Delta\hat{\bm\sigma}_{12}^\perp\right)\delta(t - t^{\rm coll}_{12})\,,
    \label{eq: force}
\end{equation}
where $\delta(t- t^{\rm coll}_{12})$ is the Dirac delta function at collision time $t^{\rm coll}_{12}$. 
The force $\mathbf F_{12}$ can be considered as an impulsive odd interaction, which typically governs the dynamics of chiral active particles.
Indeed, $\mathbf F_{12}$ is nonconservative because it cannot be derived from a potential and has a component acting transversely to the direction connecting the particles' centers~\cite{caporusso2024phase,caprini2025Bubble,maire2025hyperuniformity}.
Such odd interactions are coarse-grained descriptions of transverse forces generated by rotation, either through hydrodynamic coupling between rotating objects in a fluid~\cite{tan2022odd, massana2021arrested, mecke2023simultaneous} or through rotational friction in granular spinners. Motivated by the latter, we also include dissipative collisions by taking $\alpha<1$.

\begin{figure}[t]
    \centering
    \includegraphics[width=1\linewidth]{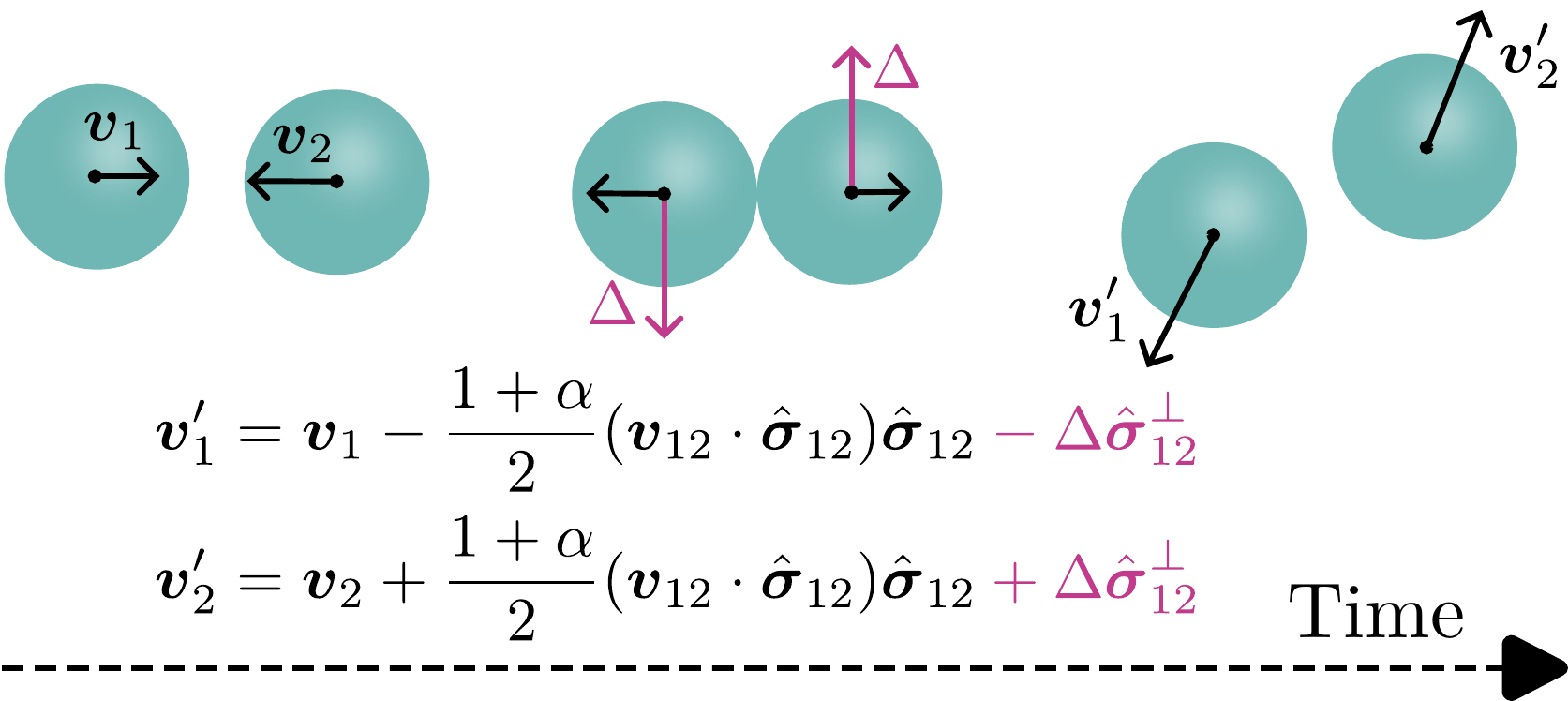}
    \caption{
    Illustration of a typical collision between two chiral active particles, labeled $1$ and $2$, following the collision rule~\eqref{eq: collision rule}. The post-collisional velocities $\bm v'_1$ and $\bm v'_2$ are characterized by (i) a reduced \textit{normal} magnitude due to dissipation, as in conventional granular particles ($\alpha < 1$), and (ii) an additional velocity component transverse to the line connecting the particle centers, generated by chirality.
    }
    \label{fig: collision}
\end{figure}

In contrast to other chiral models that include self-rotation~\cite{eren2025collisional, lubensky2005phenomenological, tsai2005chiral, han2021fluctuating, nguyen2014emergent, van2016spatiotemporal, yeo2015collective, furthauer2012active, gao2025liquidgascriticalityhyperuniformfluids}, or self-propulsion~\cite{van2008dynamics, sevilla2016diffusion, kuroda2023microscopic, kuroda2025singular, caprini2024self, lei2019nonequilibrium}, our model has neither rotational nor orientational degrees of freedom. Instead, chirality emerges solely from the injection of orbital angular momentum during collisions (Fig.~\ref{fig: collision}). This may be the simplest model for chiral systems, where collisions inject angular momentum and energy via $\Delta$, which is dissipated through the restitution coefficient $\alpha$, leading to a nonequilibrium steady state. 

We simulate the system through event-driven molecular dynamics simulations~\cite{smallenburg2022efficient}. Since the only microscopic dimensional parameters are $\sigma$, $m$, and $\Delta$, velocities are measured in units of $|\Delta|$, energies such as kinetic temperature $T=\frac{m}{2N}\sum_i \bm v_i^{2}$ in units of $m\Delta^{2}$, and times in units of $\sigma/|\Delta|$. We thus fix the transverse velocity injection $\Delta$ and use the packing fraction $\phi=N\pi\sigma^{2}/4L^{2}=n\pi\sigma^2/4$ and restitution coefficient $\alpha$ as control parameters. A strict equilibrium limit is unattainable at fixed $\Delta$. As $\alpha \to 1$, the normal component of collisions becomes elastic and non-dissipative, while collisions continue to inject energy and angular momentum: as a consequence, the steady-state kinetic temperature grows without bound. In this limit, the chiral energy injected per collision becomes asymptotically negligible. Indeed, we will show that $T \simeq m\Delta^{2}/(1-\alpha^{2})$, so that $m\Delta^{2}/T \sim 1-\alpha^{2} \ll 1$ as $\alpha\to 1$. Most chiral effects are therefore expected to be most pronounced at small $\alpha$, where $T\simeq m\Delta^{2}$, and at large $\phi$, since collisions are the source of chirality.

This chirality gives rise to new macroscopic phenomena, including odd transport coefficients and a nonzero torque density. We now describe these effects and summarize our main results.

\section{Results overview}\label{sec: results overview}

The many-body dynamics resulting from the kinetic model with collision rule~\eqref{eq: collision rule} is governed by hydrodynamic equations for the slow fields at large scales~\cite{hansen2013theory}. Number density $n(\mathbf r, t)$ and momentum $mn(\mathbf r, t) \bm u(\mathbf r, t)$ are locally conserved, so their relaxation times are set by the observation scale. While the kinetic temperature $T(\mathbf r, t)$ is not conserved for $\alpha<1$ or $\Delta\neq0$, it remains a slow field close to $\alpha=1$ and $\Delta=0$, and should therefore be included in the hydrodynamic description~\cite{dufty2011choosing, brilliantov2010kinetic}.

We obtain the following hydrodynamic balance equations at coarse-grained scales:
\begin{subequations}
    \begin{flalign}
    \label{eq:continuityEq}
        \partial_t n + \bm u \cdot \bm \nabla n &= -n\bm\nabla \cdot \bm{u}\,, \\
    \label{eq:momentumEq}
        \partial_t \bm{u} + \bm{u}\cdot\bm \nabla \bm{u}&= \dfrac{1}{mn}\bm \nabla\cdot \left(\bm \Pi^{\rm homo}+\bm \Pi^{\rm visc}\right), \\
    \label{eq:temperatureEq}
        \partial_t T + \bm{u}\cdot\bm \nabla T &= \dfrac{1}{n}\left(\bm{\Pi}:\bm \nabla \bm{u}-\bm \nabla \cdot \bm J\right) + \delta\dot{ T} \,,
    \end{flalign}
    \label{eq: hydro intro}
\end{subequations}
where $\bm \Pi=\bm \Pi^{\rm homo}+\bm \Pi^{\rm visc}$ denotes the stress tensor, decomposed into homogeneous and viscous contributions. The quantity $\bm J$ represents the heat current, while $\delta\dot T$ denotes the rate of energy change due to collisions.

\begin{figure}
    \centering
    \includegraphics[width=0.95\linewidth]{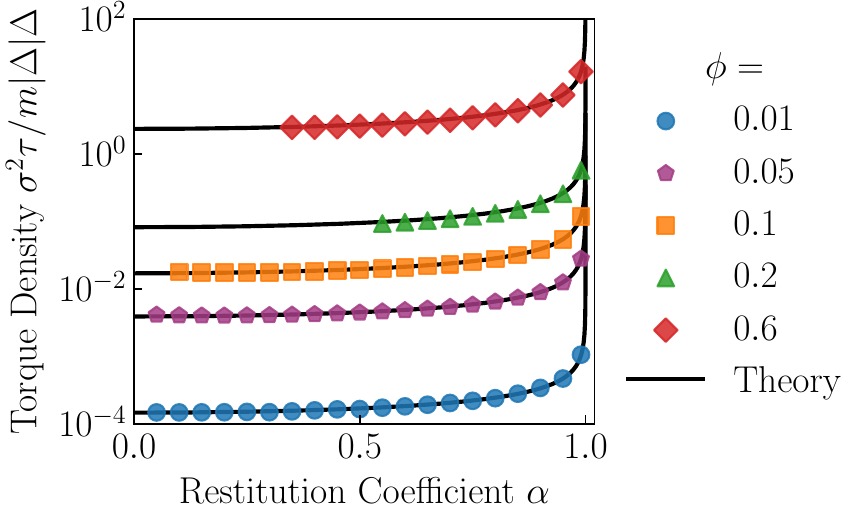}
    \caption{
    Comparison between numerical measurements and theoretical predictions [Eq.~\eqref{eq: theo tau}] for the torque density $\tau$ as a function of the restitution coefficient $\alpha$. Results are obtained for $N = 10^4$ particles, with each data point averaged over at least $10^3$ independent snapshots. Here, we include only data showing a homogeneous configuration. Therefore, numerical measurements at low $\alpha$ are not included since the system exhibits an inhomogeneous phase reminiscent of the bubble phase numerically observed in Refs.~\onlinecite{caprini2025Bubble, shen2023collective, digregorio2025phase, guo2025chirality}.
    }
    \label{fig: torque}
\end{figure}

As a first main result, we discover that the homogeneous stress tensor $\bm\Pi^{\rm homo}$ includes not only the hydrostatic pressure $p$ but also a torque density~\cite{chaikin1995principles}~$\tau=\bm \varepsilon: \bm\Pi^{\rm homo}/2$:
\begin{equation}
    \bm \Pi^{\rm homo}=-p\bm 1 + \tau\bm\varepsilon=\begin{pmatrix}
    -p & \tau \\
    -\tau & -p 
    \end{pmatrix}\,.
\end{equation}
This torque density is a direct consequence of the tangential forces during collisions acting as a source of angular momentum, and will be shown to be given by
\begin{equation}
    \tau = n\phi\chi m\sqrt{\dfrac{4\pi\Delta^2}{1 - \alpha^2}}\Delta\,,
    \label{eq: theo tau}
\end{equation}
where $\chi$ is the pair correlation at contact, which we approximate by its equilibrium hard-disk value~\cite{mulero2009equation}. The term $\tau$ has the same sign as the chiral parameter $\Delta$ and diverges as $\alpha\to 1$. This divergence reflects an unbounded rate of angular-momentum injection, which arises from the diverging temperature and collision frequency in the same limit. We find excellent agreement between the measured $\tau$ in our simulations and Eq.~\eqref{eq: theo tau} in Fig.~\ref{fig: torque}, even at high densities. 

\begin{figure*}
    \centering
    \includegraphics[width=0.95\linewidth]{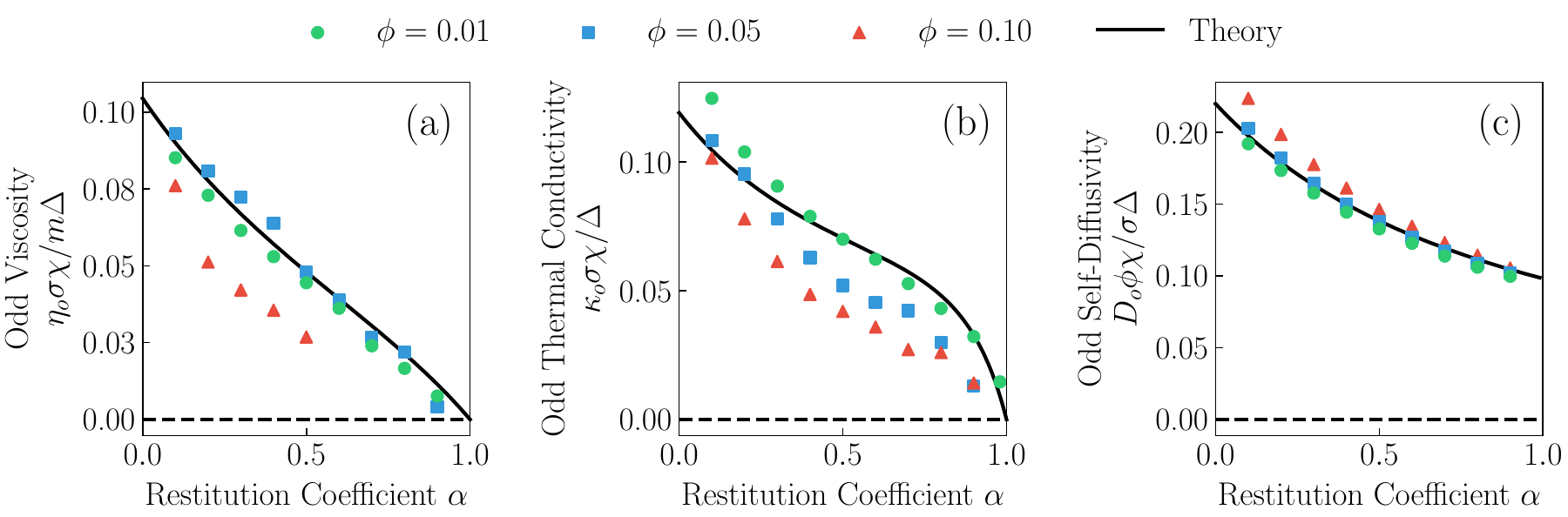}
    \caption{
    Odd transport coefficients as functions of the restitution coefficient $\alpha$—which also sets the strength of chirality---at three different packing fractions, $\phi = 0.01, 0.05, 0.1$.
    (a) Odd viscosity measured from an imposed shear flow, as detailed in Appendix~\ref{app: viscosity measurement}. The missing data points for $\phi = 0.1$ correspond to systems in which the smallest numerically accessible shear rate was still too large to yield reliable measurements.
    (b) Odd thermal conductivity measured using a Green–Kubo relation (see Appendix~\ref{app: conductivity measurement}).
    (c) Odd self-diffusivity measured using a Green–Kubo relation (see Appendix~\ref{sec: diffusivity measurement}).
    Solid lines show theoretical predictions [Eqs.~\eqref{eq: 1}, \eqref{eq: 2}, and \eqref{eq: 3}], while dashed lines indicate the corresponding values for a nonchiral system.
    }
    \label{fig: transport odd}
\end{figure*}

Our second main result is the analytical prediction for regular and odd transport coefficients~\cite{banerjee2017odd, fruchart2023odd} in the dilute limit via a Chapman-Enskog expansion. Specifically, the viscous stress
\begin{equation}
    \Pi^{\rm visc}_{ij}\equiv \eta_{ijkl}\partial_k u_l\,,
    \label{eq: viscosity from pressure}
\end{equation} 
depends on the velocity gradient and the viscosity tensor $\bm{\eta}$, with repeated indices implicitly summed. In a two-dimensional, isotropic but parity-breaking fluid, $\eta_{ijkl}$ can be decomposed into a basis of 6 independent isotropic tensors built from $\delta_{ij}$ and $\varepsilon_{ij}$. A physically motivated decomposition is~\cite{fruchart2023odd}:
\begin{equation}
    \begin{split}
        \eta_{ijkl}&= \eta_s(\delta_{i k}\delta_{j l}+\delta_{il}\delta_{jk}-\delta_{ij}\delta_{kl})+
        \zeta\delta_{ij}\delta_{kl}\\
        &+\eta_o(\varepsilon_{ik}\delta_{jl}+\varepsilon_{j l}\delta_{ik})-\eta_B\delta_{ij}\varepsilon_{kl}\\&-\eta_A\varepsilon_{ij}\delta_{kl}+\eta_R\varepsilon_{ij}\varepsilon_{kl}\,,
    \end{split}
    \label{eq: viscosity decomposition}
\end{equation}
where $\eta_s$ and $\eta_o$ denote the even and odd shear viscosities, respectively. 
Odd viscosity, in contrast to the (even) shear viscosity, is nondissipative: the odd-viscous stress is orthogonal to the velocity gradient and therefore does not produce viscous heating, $\Pi^{\rm odd}_{ij}\partial_i u_j=0$. In simple shear, $u_x=\dot\gamma y$, it yields an anisotropic, pressure-like response rather than a frictional shear stress~\cite{lapa2014swimming}.
Within the dilute approximation considered here, all viscosity coefficients other than $\eta_o$ and $\eta_s$ vanish in our system, and, remarkably, we are able to report one of the first analytical expressions for $\eta_o$ as an explicit function of the model parameters.
\begin{equation}
    \eta_o=\frac{2 m(1-\alpha)}{\chi\sigma(1 + \alpha)P(\alpha)}\Delta \,,
    \label{eq: 1}
\end{equation} 
where $P(\alpha)$ is a positive polynomial defined in Appendix~\ref{sec: viscosity}. As expected, $\eta_o$ changes sign with the chiral parameter $\Delta$ and vanishes when $\Delta\to 0$ but also as $\alpha\to 1$, showing that dissipative collisions are required to generate odd viscosity in our model. Measurements of $\eta_o$ are compared with theory in Fig.~\ref{fig: transport odd}(a) and show good agreement at low density.

The temperature expression in Eqs.~\eqref{eq: hydro intro} reflects the balance between mechanical work, heat transport, and active energy injection. The heat current is:
\begin{equation}
    \bm J=-\kappa\bm\nabla T-\kappa_o(\bm\varepsilon\cdot\bm\nabla)T\,,
\end{equation}
where $\kappa$ is the regular thermal conductivity and $\kappa_o$, its parity-odd counterpart~\cite{eren2025collisional, fruchart2022odd}. Unlike odd viscosity, odd thermal conductivity is subtler because $\bm\nabla\cdot(\kappa_o\bm\varepsilon\cdot\bm\nabla T)= (\bm\nabla\kappa_o)\cdot(\bm\varepsilon\cdot\bm\nabla T)$ since $\varepsilon_{ij}\partial_i\partial_jT=0$. Hence $\kappa_o$ does not enter the \emph{linearized bulk} temperature dynamics about a uniform state, but it affects boundary heat currents and contributes nonlinearly when $\kappa_o(n, T)$ varies in space. Within our theory, we find the value of $\kappa_o$ to be:
\begin{equation}
    \kappa_o=\frac{8  \left(1-\alpha \right)}{\chi \sigma Q(\alpha)}\Delta\,,
    \label{eq: 2}
\end{equation}
with $Q(\alpha)$ a positive polynomial defined in Appendix~\ref{sec: conductivity}. Once again, this odd coefficient vanishes in the limit $\alpha\to 1$. Overall, the theory and simulations agree well, as seen in Fig.~\ref{fig: transport odd}(b), with the low-density discrepancies most likely due to systematic measurement errors as will be explained later.

Before deriving these results, we briefly comment on transport coefficients that do not enter the hydrodynamic equations [Eqs.~\eqref{eq: hydro intro}]. Since the fluid velocity is conserved and density is governed solely by the continuity equation with no diffusive flux, there is no mass-diffusion mode in the hydrodynamic equations. Nonetheless, treating a fluid particle as a tracer in the effective bath formed by the other particles leads to an effective diffusion equation for the tracer density, with current
\begin{equation}
\bm J_s=-D\bm\nabla n_s-D_o(\bm\varepsilon\cdot\bm\nabla)n_s\,,
\end{equation}
where $D$ and $D_o$ are the regular and odd self-diffusivities~\cite{hargus2021odd, caprini2025active}, respectively. As for the odd thermal conductivity, the effect of the odd diffusivity primarily enters through boundary conditions. We predict $D_o$ to be given by
\begin{equation}
    D_o=\dfrac{\pi\sigma}{2\phi\chi (1+\alpha)R(\alpha)}\Delta\,,
    \label{eq: 3}
\end{equation}
with $R(\alpha)$ a positive polynomial defined in Appendix~\ref{sec: diffusivity}. This coefficient does not vanish in the limit $\alpha\to 1$ and is very well predicted by the theory, as shown in Fig.~\ref{fig: transport odd}(c). 

All other regular transport coefficients, as well as the pressure and temperature, are also well captured by our theory and are presented later in the article. We now explain how these expressions are derived.

\section{Results derivation}\label{sec: results derivation}
All analytical derivations and integral evaluations appearing implicitly or explicitly in this section were performed symbolically using SymPy~\cite{sympy} in a notebook provided in the Supplementary Material (SM).

\subsection{Boltzmann equation}

\begin{figure*}
    \centering
    \includegraphics[width=0.8\linewidth]{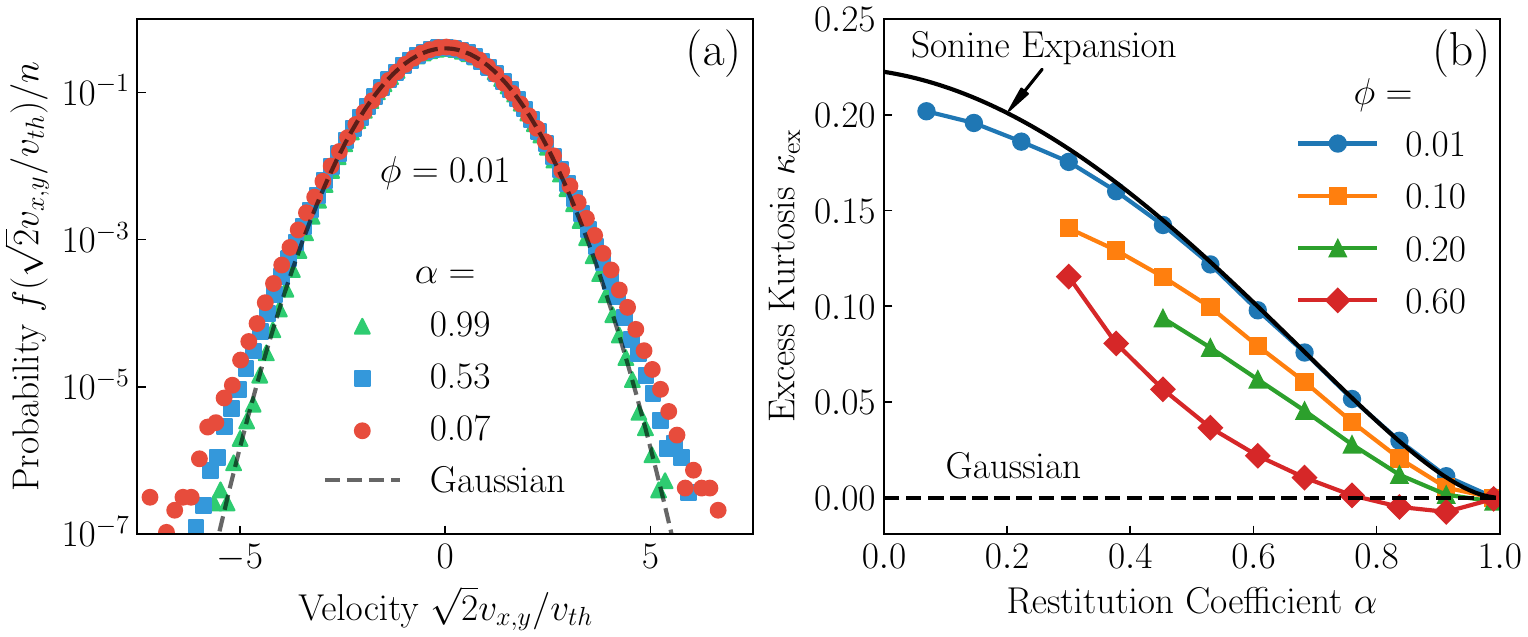}
    \caption{
    Comparison of the steady-state velocity probability distribution $f(v)$ between simulations ($N = 5000$) and theory.
    (a) Distribution of the normalized velocity in a dilute system ($\phi = 0.01$) for several values of the restitution coefficient $\alpha$. We use $v_{\mathrm{th}} = \sqrt{2T/m}$ so that the distributions have unit variance.
    (b) Excess kurtosis $\kappa_{\rm ex}$ as a function of $\alpha$ at different packing fractions. Solid and dashed lines show theoretical predictions: $\kappa_{\rm ex} = 0$ corresponds to a Gaussian distribution, while a Sonine expansion predicts $\kappa_{\rm ex} \neq 0$ in the dilute limit where molecular chaos holds. Data at low $\alpha$ are omitted for some values of $\phi$ when the system develops inhomogeneous configurations, reminiscent of those observed in Refs.~\onlinecite{caprini2025Bubble,shen2023collective, digregorio2025phase, guo2025chirality}.}
    \label{fig: distrib}
\end{figure*}

Because collisions are instantaneous and strictly binary, the microscopic dynamics in Eqs.~\eqref{eq: collision rule} admit an approximate description in terms of a Boltzmann–Enskog equation~\cite{brilliantov2010kinetic, chapman1990mathematical}
\begin{equation}
    \partial_t f(\mathbf r_1, \bm v_1, t)  + \bm v_1\cdot \bm\nabla f(\mathbf r_1, \bm v_1, t)= {\mathcal J}\big(\mathbf r_1, \bm v_1|f, f\big)\,,
    \label{eq: boltzmann equation}
\end{equation}
where $f$ is the single-particle distribution for observing a particle at time $t$ with position $\mathbf{r}_1$ and velocity $\mathbf{v}_1$. The term ${\mathcal J}$ represents the collision operator, accounting for the velocity change at a given position due to a collision with another particle. 
This operator can be obtained by counting the number of collisions per unit of time at $(\mathbf r_1, \bm v_1)$ in phase space, i.e.\ by calculating $\nu(\mathbf r_1,\bm v_1,\bm v_2,\hat{\bm\sigma}_{12})d \mathbf r_1 d\bm v_1 d\bm v_2d\hat{\bm\sigma}_{12}$, where $\nu$ is the differential collision frequency, proportional to the relative speed and the scattering cross section, encoding the probability of a binary collision with impact direction $\hat{\bm\sigma}_{12}=(\mathbf{r}_1-\mathbf{r}_2)/\sigma$.
For hard-disks~\cite{brilliantov2010kinetic}, the collision frequency can be expressed as $\nu(\mathbf r_1,\bm v_1,\bm v_2,\hat{\bm\sigma}_{12})= f_2(\mathbf r_1, \mathbf r_2, \bm v_1, \bm v_2) \sigma |\bm {v}_{12}\cdot \hat{ \bm\sigma}_{12}|$, with $f_2$ the two-particle density distribution function calculated at the coordinate of the second particle, $\mathbf r_2=\mathbf r_1 - \sigma \hat{\bm\sigma}_{12}$ and $\bm{v}_{12}=\bm{v}_1-\bm{v}_2$.
Collisions can remove particles from velocity $\bm v_1$ (loss term) or bring particles from some pre-collisional velocity $\bm v_1''$ into $\bm v_1$ (gain term):
    \begin{flalign}
    \tilde{\mathcal J}&(\mathbf r_1, \bm {v}_1) =-\!\!\int \Theta(-\bm {v}_{12}\cdot \hat{ \bm\sigma}_{12} )\nu(\mathbf r_1,\bm v_1,\bm v_2,\hat{\bm\sigma}_{12})d\bm v_2d\hat{\bm\sigma}_{12}\nonumber\\
    & +\!\!\int \Theta(-\bm {v}_{12}''\cdot \hat{ \bm\sigma}_{12} )\nu(\mathbf r_1,\bm v_1'',\bm v_2'',\hat{\bm\sigma}_{12})
    d\bm v_2''d\hat{\bm\sigma}_{12}\,,
    \label{eq: J  intermediate}
\end{flalign}
with $\Theta$ the Heaviside function ensuring head-on collisions and $\bm v''$ the pre-collisional velocities that produce $\bm v$ after a collision. Inverting the collision rule (Eqs.~\eqref{eq: collision rule}) yields:
\begin{subequations}
\label{eq:operator}
    \begin{flalign}
        \bm v_1''&= \bm v_1 - \dfrac{1+\alpha^{-1}}{2}(\bm v_{12}\cdot \hat{\bm\sigma}_{12})\hat{\bm\sigma}_{12} + \Delta \hat{\bm\sigma}_{12}^\perp\,, \\
        \bm v_2''&= \bm v_2 + \dfrac{1+\alpha^{-1}}{2}(\bm v_{12}\cdot \hat{\bm\sigma}_{12})\hat{\bm\sigma}_{12} - \Delta\hat{\bm\sigma}_{12}^\perp\,.
    \end{flalign}
\end{subequations}
Since $\nu$ depends on the two-particle distribution density function, the evolution equation for $f$ is not closed. To proceed, we adopt the molecular chaos closure, which neglects velocity-velocity and velocity-position correlations at collision~\cite{pagonabarraga2001randomly}:
\begin{equation}
    f_2(\mathbf r_1, \mathbf r_2, \bm v_1, \bm v_2)=\chi(\mathbf r_1, \mathbf r_2)f(\mathbf r_1, \bm v_1) f(\mathbf r_2, \bm v_2)\,,
    \label{eq: molecular chaos}
\end{equation}
where $\chi(\mathbf r_1, \mathbf r_2)\equiv\chi\big[n\big((\mathbf r_1+\mathbf r_2)/2\big)\big]$ corresponds to the pair correlation function $\chi[n]$ at contact evaluated at the local midpoint density~\cite{chapman1990mathematical}, taking into account short–range spatial correlations arising from the finite size of the particles.

Putting everything together and changing variables $\bm v''\to \bm v$ in the second term of Eq.~\eqref{eq: J  intermediate}, we obtain the collision operator of the Boltzmann-Enskog equation~\cite{brilliantov2010kinetic}:
\begin{equation}
    \begin{split}
    &{\mathcal J}(\mathbf r_1, \bm {v}_1|f, f) = \sigma \int \Theta(-\bm {v}_{12}\cdot \hat{ \bm\sigma}_{12} )|\bm {v}_{12}\cdot \hat{ \bm\sigma}_{12}|\chi(\mathbf r_1, \mathbf r_2)\\
    &\times \Biggl[\dfrac{f(\mathbf r_1, \bm v_1'')f(\mathbf r_2, \bm v_2'')}{\alpha^2}
    -f(\mathbf r_1, \bm v_1)f(\mathbf r_2, \bm v_2)\Biggr]d \bm v_2d\hat{\bm\sigma}_{12} \,.
    \end{split}
    \label{eq: J dense}
\end{equation}
Here, the prefactor $1/\alpha^{2}$ in the gain term comes from two contributions: (i) the Jacobian of the linear change of variables $(\bm v_1'',\bm v_2'')\mapsto(\bm v_1,\bm v_2)$, which yields $d\bm v_2''=(1/\alpha)d\bm v_2$, and
(ii) the transformation of the collision rate, which involves the normal relative velocity
$|\bm v_{12}''\cdot\hat{\bm\sigma}_{12}|=(1/\alpha)|\bm v_{12}\cdot\hat{\bm\sigma}_{12}|$ for the inelastic rule. Neither the Jacobian nor the normal-rate factor are affected by the chiral parameter $\Delta$, since it induces only an \textit{additive} and \textit{transverse} change of the velocities.
The shorthand ${\mathcal J}(\mathbf r_1,\bm v_1|f,f)$ highlights the bilinear dependence of the collision operator on the single-particle distributions of the two colliding particles, $f(\mathbf r_1,\bm v_1)$ and $f(\mathbf r_2,\bm v_2)$. We keep this bilinear structure explicit, since later we will also consider mixed terms of the form ${\mathcal J}(\mathbf r_1,\bm v_1|f,g)$, where $g$ is a function distinct from $f$.

Eq.~\eqref{eq: boltzmann equation} is a nonlinear and nonlocal integro-differential equation and is therefore difficult to analyze. We first familiarize ourselves with it by considering the spatially homogeneous state.

\subsection{Homogeneous probability distribution}
\label{sec: proba}

Our objective is to derive hydrodynamic equations from the microscopic collision rule (Eq.~\eqref{eq: collision rule}). Before coarse-graining, we note that non-gradient observables, such as the homogeneous stress, are already accessible in the spatially uniform steady state, where time and space derivatives vanish. Consequently, Eq.~\eqref{eq: boltzmann equation} reduces to:
\begin{equation}
    {\mathcal J}(\bm v| f, f) = 0\,,
    \label{eq: solution to f}
\end{equation}
which determines the steady-state single-particle distribution $f(\mathbf r, \bm v, t)\equiv f(\bm v, t)$. In the equilibrium limit ($\alpha\to1$ and $\Delta\to0$), detailed balance yields a Maxwellian shape, whereas here dissipation and parity-breaking may drive a nontrivial steady state.

We first assess whether non-Gaussianities are quantitatively important by simulating the microscopic dynamics defined by the collision rule~\eqref{eq: collision rule}. Fig.~\ref{fig: distrib}(a) shows the steady-state velocity distribution in the dilute regime. At weak normal dissipation ($\alpha=0.99$), the velocity distribution $f(\bm v)$ is effectively Gaussian. As shown in Appendix~\ref{sec: sonine}, this behavior stems from the divergence of the temperature as $\alpha \to 1$, which makes the nonequilibrium chiral contribution asymptotically negligible. In this limit, $m\Delta^{2} \ll T$, and the system approaches an equilibrium-like regime. As dissipation $(1-\alpha)$ increases, chiral effects become more prominent relative to the temperature $T$ ($m\Delta^2\simeq T$), leading to small but systematic overpopulated tails. In the dilute limit $\phi \to 0$, these tails can be treated perturbatively by expanding $f$ in Sonine polynomials, which form an orthogonal basis with respect to the Gaussian weight~\cite{brilliantov2010kinetic}. The derivation of this expansion beyond the Gaussian case is reported in Appendix~\ref{sec: sonine}. The results of this expansion are presented in Fig.~\ref{fig: distrib}(b) via the 2D excess kurtosis $\kappa_{\rm ex} \equiv \langle \bm v^4\rangle/\langle \bm v^2\rangle^2 - 2$, which quantifies deviations from Gaussianity since $\kappa_{\rm ex}=0$ for a Gaussian distribution. The Sonine prediction is in good agreement with simulations at low packing fractions $\phi$, independently of the restitution coefficient $\alpha$. However, its accuracy deteriorates at higher densities, where velocity correlations at collisions invalidate the molecular-chaos assumption~\cite{pagonabarraga2001randomly} (see Eq.~\eqref{eq: molecular chaos}). A Sonine expansion cannot capture the resulting non-Gaussian features of the velocity distribution, which are notoriously difficult to treat analytically~\cite{soto2001Precollisional, van1998ring}. Since Sonine corrections are quantitatively small in the regime of interest, we neglect them in the following. They can be straightforwardly reintroduced, at the expense of more cumbersome expressions.

We note that chiral hard-particle models with internal spin can exhibit pronounced non-Gaussian statistics~\cite{eren2025collisional}, which can limit quantitative accuracy. Since chirality in our model arises without internal rotation, non-Gaussian corrections remain small.

\begin{figure*}[t]
    \centering
    \includegraphics[width=0.9\linewidth]{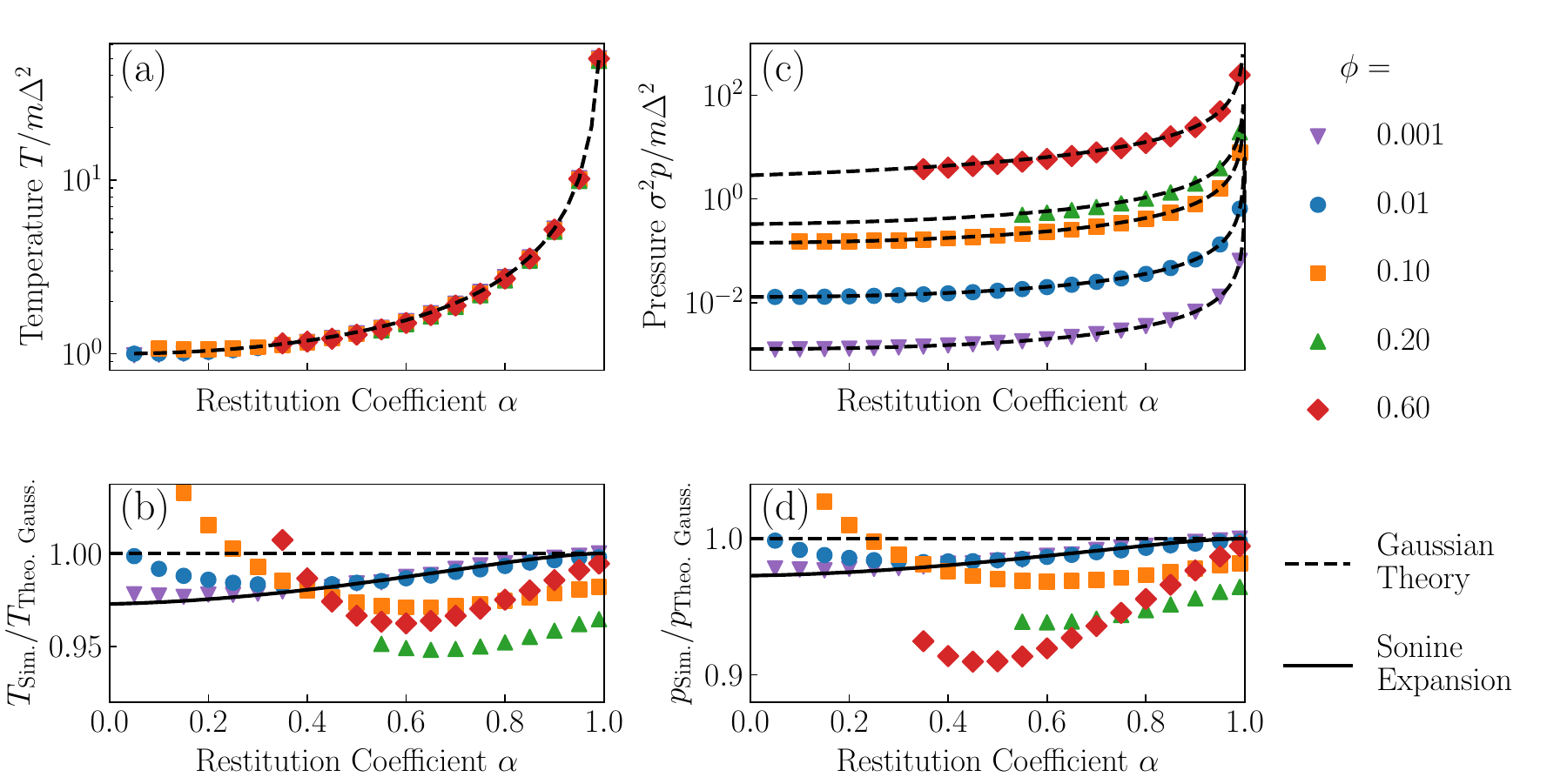}
    \caption{
    Homogeneous temperature $T$ [(a),(b)] and pressure $p$ [(c),(d)] as functions of the restitution coefficient $\alpha$ for different values of the packing fraction $\phi$. Panels (a) and (c) show the raw quantities, while panels (b) and (d) display the same data normalized by their Gaussian theoretical predictions (Eq.~\eqref{eq: temperature iso} and Eq.~\eqref{eq: pressure iso} with $T$ replaced by its Gaussian prediction). Results are obtained for $N = 10^4$ particles, with each data point averaged over at least $10^3$ independent snapshots. Only data corresponding to homogeneous configurations are shown; data exhibiting inhomogeneous states---typically occurring at large $\phi$ and low $\alpha$ and reminiscent of the nonuniform phase observed in Refs.~\onlinecite{caprini2025Bubble,shen2023collective, digregorio2025phase, guo2025chirality}---are omitted.
    }
    \label{fig: iso}
\end{figure*}

\subsection{Homogeneous temperature and stress}

The homogeneous temperature is defined through the velocity fluctuations as:
\begin{equation}
    T = \dfrac{m}{2n}\int \bm v^2f(\bm v)d\bm v\,.
\end{equation}
Since we consider a Gaussian approximation $f(\bm v)$, motivated by Fig.~\ref{fig: distrib}(a),
\begin{equation}
    f(\bm v)=\dfrac{mn}{2\pi T}e^{\displaystyle -\dfrac{m\bm v^2}{2T}}\,,
\end{equation}
the temperature field can be identified with the variance of the distribution.
Multiplying Eq.~\eqref{eq: boltzmann equation} by $m\bm v^2$ and integrating over $\bm v$ yields an evolution equation for the temperature~\cite{maire2024interplay, brito2013hydrodynamic} (see Appendix.~\ref{sec:balance})
\begin{equation}
    \partial_t T(t)=\dfrac{\omega(\phi, T)}{2}\langle \delta E\rangle_{\rm coll}\equiv \delta \dot T\,,
    \label{eq: temperature}
\end{equation}
where $\delta E\equiv m(\bm v_1'^2+\bm v_2'^2-\bm v_1^2-\bm v_2^2)/2$ is the energy change at collision and $\omega$ the collision frequency per particle~\cite{pagonabarraga2001randomly}. As a consequence, the term $\omega\langle\delta E\rangle_{\rm coll}$ coincides with the average rate of energy change due to collisions and can be expressed as:
\begin{equation}
\begin{aligned}
\omega\langle\delta E\rangle_{\rm coll}=&{\sigma\chi}{n^{-1}}\int \Theta(-\bm {v}_{12}\cdot \hat{ \bm\sigma}_{12})|\bm {v}_{12}\cdot \hat{ \bm\sigma}_{12}|\\
&\times\delta E f(\bm v_2)f(\bm v_1)d\hat{\bm\sigma}_{12}d\bm v_1 d \bm v_2 \,.
\end{aligned}
\end{equation}
Imposing stationarity in Eq.~\eqref{eq: temperature} yields a theoretical prediction for the temperature (see Appendix.~\ref{sec: kinetic inte}):
\begin{equation}
   T_{\rm Gauss} = \dfrac{m\Delta^2}{1-\alpha^2}\,,
   \label{eq: temperature iso}
\end{equation}
which follows from the condition $0=\langle \delta E\rangle_{\rm coll}= m\Delta^2 - (1-\alpha^2)T$.
This steady-state temperature is independent of packing fraction because higher density increases the collision rate but not the \textit{average} energy exchanged per collision in the steady state, leaving the balance unchanged~\cite{brito2013hydrodynamic}. In practice, the breakdown of molecular chaos invalidates this argument.

Even in the limit $\alpha \to 1$ at finite density, we find $T_{\rm sim}/T_{\rm theory} \neq 1$, indicating that nonequilibrium correlations beyond molecular chaos persist, even though the velocity distribution $f$ is nearly Gaussian.  At moderate packing fractions $\phi$, small deviations from a Gaussian distribution manifest as a weak density dependence of $T$, demonstrating the breakdown of molecular chaos. Nevertheless, the temperature is overall well captured by Eq.~\eqref{eq: temperature iso}, as shown in Fig.~\ref{fig: iso}(a–b), and the agreement can be further improved by including a Sonine correction (see Appendix~\ref{sec: sonine}).

The homogeneous stress can be computed without invoking the Boltzmann equation by using the standard expression for the virial stress~\cite{hansen2013theory}:
\begin{equation}
    \bm\Pi^{\rm virial} = -\dfrac{\sum_{\alpha=1}^{N} m\bm c_\alpha \otimes \bm c_\alpha-  \sigma\sum_{\alpha<\beta}^{N}\hat{\bm\sigma}_{\alpha\beta}\otimes \mathbf F_{\alpha\beta}}{L^2}\,,
    \label{eq: virial}
\end{equation}
where $\bm c_\alpha = \bm v_\alpha - \langle \bm v \rangle$ is the deviation of the velocity of particle $\alpha$ from the local average velocity and $\mathbf F_{\alpha\beta}$ is the singular force given in Eq.~\eqref{eq: force}. In the homogeneous state, the average virial stress coincides with the homogeneous stress tensor $\langle \bm\Pi^{\rm virial}\rangle =\bm\Pi^{\rm homo}$. The interaction term arises from collisions and can be obtained from a time average over a window $\mathcal T$, which reduces to a collisional average~\cite{maire2024non, soto2001statistical}:
%\begin{equation}
    \begin{flalign}
    \big\langle \hat{\bm\sigma}_{\alpha\beta}\otimes \mathbf F_{\alpha\beta}\big\rangle
    &=\mathcal T^{-1}\int_{0}^{\mathcal T}\hat{\bm\sigma}_{\alpha\beta}(t)\otimes \mathbf F_{\alpha\beta}(t)dt\nonumber\\
    &=\mathcal T^{-1}\sum_{\rm all~\alpha\beta~coll.}\hat{\bm\sigma}_{\alpha\beta}(t^{\rm coll}_{\alpha\beta})\otimes \mathbf I_{\alpha\beta}(t^{\rm coll}_{\alpha\beta})\nonumber\\
    &=\dfrac{\omega(\phi, T)}{2}\langle \hat{\bm\sigma}_{\alpha\beta}\otimes \mathbf I_{\alpha\beta}\rangle_{\rm coll}\,.
     \label{eq: coll to coll}
    \end{flalign}
%\end{equation}
Here, we have used Eq.~\eqref{eq: force} to define: $\mathbf I_{\alpha\beta}/m=-(1+\alpha)(\bm v_{\alpha\beta}\cdot \hat{\bm\sigma}_{\alpha\beta})\hat{\bm\sigma}_{\alpha\beta}/2 -\Delta \hat{\bm\sigma}_{\alpha\beta}^\perp$.
Equations~\eqref{eq: virial} and \eqref{eq: coll to coll} yield an explicit expression for the homogeneous stress tensor $\bm\Pi^{\rm homo}$ as a function of the model parameters (see SM):
\begin{subequations}
    \begin{flalign}
    &\bm \Pi^{\rm homo}= -p\bm 1 +\tau\bm\varepsilon\,,\\
    &p=p^{\rm kinetic} + p^{\rm coll}=n T + n\phi \chi(1+\alpha)T\,,\\
    &\tau = \tau^{\rm coll}=n\phi \chi \sqrt{4\pi m T}\Delta\,.
\end{flalign}
\label{eq: pressure iso}
\end{subequations}
In this expression, the pressure reduces to the standard granular-gas result~\cite{pagonabarraga2001randomly}, since the transverse chiral impulse does not contribute to the momentum transfer along the line connecting the particle centers, $\hat{\bm \sigma}$. By contrast, chirality generates a purely collisional torque density, $\tau$, through the $\Delta$ term in the collision rule. The pressure includes an ideal-gas contribution of order $\mathcal{O}(n)$ from $m \bm c \otimes \bm c$, whereas $\tau$ is purely collisional, of order at least $\mathcal{O}(n^2)$, so that $|\tau| \ll p$ at low $\phi$.

We compare simulations and theory in Fig.~\ref{fig: iso}(c–d) by replacing $T$ with its theoretical value (Eq.~\eqref{eq: temperature iso}). The agreement is excellent, even though we use the equilibrium hard-disk estimate~\cite{mulero2009equation}$ \chi\approx\chi^{\rm eq}\simeq (1 - 7\phi/16 - \phi^3/20)/(1-\phi)^2$. An approximate Sonine correction is again proposed where the temperature in Eq.~\eqref{eq: pressure iso} is replaced by the Sonine corrected one obtained in Appendix~\ref{sec: sonine}. The comparison between numerical and theoretical results of $\tau$ (Eq.~\eqref{eq: pressure iso}) has already been shown in Fig.~\ref{fig: torque} and shows a good agreement.

\subsection{Chapman-Enskog ansatz}

The hydrodynamic equations [Eqs.~\eqref{eq: hydro intro}] can be derived by taking time derivatives of the slow fields~\cite{dorfman2021contemporary} (see Appendix~\ref{sec:balance}):
\begin{subequations}
    \begin{flalign}
        n(\mathbf r, t)&=\int f(\mathbf r, \bm v, t)d\bm v\,,\\
        \bm u(\mathbf r, t)&=\dfrac{1}{n(\mathbf r, t)}\int \bm vf(\mathbf r, \bm v, t)d\bm v\,,\\
        T(\mathbf r, t)&=\dfrac{m}{2n(\mathbf r, t)}\int (\bm v-\bm u)^2f(\mathbf r, \bm v, t)d\bm v \,.
    \end{flalign}
    \label{eqs: definition of fields}
\end{subequations}
The homogeneous solution $f(\mathbf r, \bm v, t) = f(\bm v)$ introduced above is independent of space and time and therefore cannot be used to derive hydrodynamic equations. To capture spatial and temporal variations, we construct an approximate solution of the Boltzmann equation through a suitable coarse-graining procedure using the Chapman–Enskog expansion. Originally developed to derive the Navier–Stokes equations for Hamiltonian systems~\cite{chapman1990mathematical} and subsequently extended in various directions~\cite{dorfman2021contemporary,brilliantov2010kinetic}, this approach has more recently been applied to active matter~\cite{ihle2011kinetic, gonnella2015motility, degond2010diffusion, hancock2017statistical, barbaro2012phase,bonilla2019active, marenduzzo2016introduction, pinto2025hydrodynamic, feliachi2022fluctuating, nesbitt2021uncovering}, including chiral models with simplified dynamics~\cite{fruchart2022odd}.

The Chapman-Enskog method assumes that the space and time dependence of $f$ enters only through the hydrodynamic fields and their gradients~\cite{brey2015hydrodynamics, dorfman2021contemporary}, so that we have
\begin{equation}
    f(\mathbf r, \bm v, t)\equiv f[n(\mathbf r, t), \bm u(\mathbf r, t), T(\mathbf r, t)| \bm v]\,.
\end{equation}
To proceed, we perform a gradient expansion on the Boltzmann equation [Eq.~\eqref{eq: boltzmann equation}] by introducing a bookkeeping parameter $\mu$ set back to 1 at the end of the calculation:
\begin{equation}
    f=f^{(0)}(1 + \mu \Phi^{(1)}+\dots), \qquad\bm \nabla\to\mu \bm\nabla\,.
    \label{eq: expansion}
\end{equation}
In practice, the expansion is never performed beyond $\mathcal O(\mu^1)\sim\mathcal O(\bm \nabla^1)$, because the resulting hydrodynamic equations are typically unstable~\cite{rosenau1989extending, bobylev2006instabilities}. This ansatz is expected to hold on timescales much longer than the mean time interval between collisions, so that the dynamics is governed solely by the slow fields $n$, $\bm u$, and the quasi-slow field $T$. Here, $f^{(0)}$ corresponds to the previously obtained homogeneous solution---with density and temperature promoted to spatially varying fields---yielding the dissipationless Euler terms at order $\mathcal{O}(\bm{\nabla})$. The term $\Phi^{(1)}$ captures deviations from this local equilibrium through field gradients, supplying the $\mathcal{O}(\bm{\nabla}^2)$ viscous and thermal-transport corrections that complete the chiral Navier-Stokes equations~\cite{dorfman2021contemporary}. Regarding the values of the transport coefficients, we remark that this approach is fully consistent with the linear response one proposed in Ref.~\onlinecite{eren2025collisional} (see also Ref.~\onlinecite{fruchart2022odd}). 
%However, we believe that the Chapman-Enskog approach is physically clearer, notably because the probability distribution itself is the main object of interest. 

We will work in the dilute limit by assuming collisions occur at the same position $\mathbf r_1$ for both particles, rather than at $\mathbf r_1$ and $\mathbf r_2=\mathbf r_1-\sigma\hat{\bm\sigma}_{12}$. This yields
\begin{equation}
    \begin{split}
    &{\mathcal J}^{\rm dilute}(\mathbf r_1, \bm {v}_1| f, f) = \sigma \chi \int d \bm v_2d\hat{\bm\sigma}_{12}\Theta(-\bm {v}_{12}\cdot \hat{ \bm\sigma}_{12} )\\
    &\times |\bm {v}_{12}\cdot \hat{ \bm\sigma}_{12}|\Biggl[\dfrac{f(\mathbf r_1, \bm v_1'')f(\mathbf r_1, \bm v_2'')}{\alpha^2}
    -f(\mathbf r_1, \bm v_1)f(\mathbf r_1, \bm v_2)\Biggl]\,,
    \end{split}
    \label{eq: J dilute}
\end{equation}
which has to be compared with Eq.~\eqref{eq: J dense}. 
This approximation neglects Enskog (collisional-transfer) contributions to the fluxes, retaining only the kinetic terms~\cite{garzo2018enskog, mendez2026transportpropertiesmodelconfined}. As a result, the stress reduces to its kinetic part and is symmetric by construction (see also Appendix~\ref{sec:balance}). It therefore cannot capture $\mathcal{O}(n^{2})$ effects, such as the collisional torque density or any antisymmetric stress:
\begin{equation}
\bm \Pi = \bm \Pi^{\rm kin} = -m \int \bm c \otimes \bm c  f(\mathbf r, \bm v, t)  d\bm c \,,
\label{eq: kinetic def of stress tensor}
\end{equation}
where $\bm c = \bm v - \bm u$ and $\Pi^{\rm kin}_{ij} = \Pi^{\rm kin}_{ji}$. While odd viscosity is included---since it contributes to the symmetric part of the stress---other parity-odd viscosities that produce a nonsymmetric stress cannot be captured.

Finally, due to momentum conservation, the transport coefficients in our two-dimensional fluid exhibit a logarithmic divergence with system size~\cite{nakano2025looking,timeResibois1975}. This effect is smaller than the numerical accuracy for the system sizes considered and is therefore neglected, as it lies beyond the scope of our non-fluctuating Chapman–Enskog approach, which yields the ‘‘bare'' transport coefficients.

\subsection{Order \texorpdfstring{$\mu^0$}{0}}

At order $\mu^0$, the Boltzmann equation [Eq.~\eqref{eq: boltzmann equation}] reduces to:
\begin{equation}
    \partial_t^{(0)}f^{(0)}={\mathcal J}^{\rm dilute}(f^{(0)}, f^{(0)})\,,
\end{equation}
where $\partial_t^{(0)}$ denotes the $\mathcal O(\mu^0)$ part of the time derivative. Recalling that $f$ is a function of the hydrodynamic fields, we obtain
\begin{equation}
    \partial_tf^{(0)} = (\partial_tn)\partial_nf^{(0)} + (\partial_t\bm u)\cdot\partial_{\bm u} f + (\partial_tT)\partial_Tf\,.
\end{equation}
From Eqs.~\eqref{eq: hydro intro}, $\partial_t n$ and $\partial_t \bm u$ are $\mathcal{O}(\mu)$, as they involve gradients, whereas $\partial_t T = \delta \dot T + \mathcal{O}(\mu^1)$ is generally of order $\mathcal{O}(\mu^0)$. We assume, however, that collisions rapidly relax temperature fluctuations, so that the temperature is locally steady, $\delta \dot T(\mathbf r) \simeq 0$, making $T$ a fast field. Nonetheless, imposing $\partial_t T = 0$ at all orders, leads to unphysical values of the viscosities. For this reason, we retain $T$ as an independent hydrodynamic field with its own evolution equation, as is standard in granular hydrodynamics~\cite{dufty2011choosing}.
This approximation leads to $\partial_t^{(0)}f^{(0)}=0$ and therefore
\begin{equation}
    {\mathcal J}^{\rm dilute}(f^{(0)}, f^{(0)})=0\,.
\end{equation}
This equation is formally identical to the homogeneous one [Eq.~\eqref{eq: solution to f}] and therefore admits the same solution. In particular, we have seen that a Gaussian provides an accurate approximation. However, the parameters entering $f^{(0)}$ do not necessarily coincide \textit{a priori} with the hydrodynamic fields $n$, $T$, and $\bm u$, since these fields are defined by Eqs.~\eqref{eqs: definition of fields} and can be influenced by $\Phi^{(1)}$. Nonetheless, the Chapman-Enskog approach assumes that $f^{(0)}$ is written in terms of the hydrodynamic fields themselves~\footnote{This choice is not merely conventional: in the Chapman-Enskog expansion one \emph{identifies} the hydrodynamic fields with the corresponding velocity moments of $f^{(0)}$. As a result, the higher-order corrections do not contribute to those moments:
$\int f^{(0)}(\bm v)\Phi^{(1)}(\bm v)\Psi(\bm v)d\bm v=0$ with $\Psi\in\{1,\bm v,\bm v^{2}\}$. These moment constraints are known as the solvability (Fredholm) conditions for the linear equation $\mathcal L[\Phi^{(1)}]=S$ (defined below), which impose the source term $S$ to be orthogonal to $\ker(\mathcal L^{\dagger})$, which is spanned by the collision invariants ($\bm v^2$ is also included in our energy-nonconserving system to recover the proper nonchiral limit). Since $\mathcal L$ has a nontrivial null space, it is not invertible on the full function space. Enforcing the above orthogonality removes the null modes and guarantees the existence of $\Phi^{(1)}$.}\textsuperscript{,}\cite{bardos1991fluid}:
\begin{equation}
f^{(0)}(\mathbf r, \bm v, t)=\dfrac{mn(\mathbf r,t)}{2\pi T(\mathbf r,t)}
\exp\left[-\dfrac{m\big(\bm v-\bm u(\mathbf r,t)\big)^2}{2T(\mathbf r,t)}\right]\,,
\label{eq: f0}
\end{equation}
with the parameters promoted to slowly varying fields. One may also include the previously discussed Sonine corrections to $f^{(0)}$, but their contribution is negligible.

\subsection{Order \texorpdfstring{$\mu^1$}{1}: generality}

At order $\mu^1$, Eq.~\eqref{eq: boltzmann equation} becomes~\cite{dorfman2021contemporary}:
\begin{equation}
    \big(\partial_t^{(1)} + \bm v\cdot\bm\nabla\big) f^{(0)}\equiv\mathcal L[\Phi^{(1)}]\,,
    \label{eq: linearized boltzmann equation}
\end{equation}
where we defined $\mathcal L$ as the linear Boltzmann operator
\begin{equation}
    \mathcal L[\Phi^{(1)}]={\mathcal J}^{\rm dilute}(f^{(0)}, f^{(0)}\Phi^{(1)}) + {\mathcal J}^{\rm dilute}(f^{(0)}\Phi^{(1)}, f^{(0)})\,.
\end{equation}
Using the $\mathcal O(\mu)$ hydrodynamic equations, we have
\begin{equation}
\begin{split}
    \partial_t^{(1)} n&=-\bm\nabla\cdot(n\bm u),\\
    \partial_t^{(1)} \bm u&=-(\bm u\cdot\bm\nabla)\bm u + \dfrac{1}{mn}\bm \nabla\cdot \bm\Pi^{\rm homo},\\
    \partial_t^{(1)} T&=-\bm u\cdot\bm\nabla T - T\bm \nabla\cdot \bm u\,.
\end{split}
\label{eq: expansion time derivative order 1}
\end{equation}
In the dilute approximation, we neglect collisional transfer, hence also the torque, and use $\bm \Pi^{\rm homo}=-p\bm 1=-nT\bm 1$. Inserting Eq.~\eqref{eq: f0} into Eq.~\eqref{eq: linearized boltzmann equation} yields~\cite{brilliantov2010kinetic}
\begin{equation}
    \big(\partial_t^{(1)} + \bm v\cdot\bm\nabla\big) f^{(0)}=\dfrac{1}{T}\left(\mathbf D(\bm c):\bm \nabla \bm u +\mathbf A(\bm c)\cdot\bm \nabla \log T\right)f^{(0)}\,,
    \label{eq: expanding}
\end{equation}
where $\bm c=\bm v-\bm u$ and
\begin{equation}
\begin{split}
        \mathbf A(\bm c)&=\left(\dfrac{m\bm c^2}{2} - 2T\right)\bm c\,,\\
        \mathbf D(\bm c)&=m\left(\bm c \otimes\bm c - \dfrac{\bm c^2}{2}\bm 1\right)\,.
\end{split}
\end{equation}
Substituting Eqs~\eqref{eq: expanding} into Eq.~\eqref{eq: linearized boltzmann equation}, we obtain the equation for $\Phi^{(1)}$:
\begin{equation}
    \dfrac{f^{(0)}}{T}\big(\mathbf D(\bm c):\bm \nabla \bm u + \mathbf A(\bm c)\cdot \bm \nabla \log T\big)=\mathcal L[\Phi^{(1)}]\,.
    \label{eq: to solve}
\end{equation}
A formal solution of Eq.~\eqref{eq: to solve} requires a (pseudo-) inverse~\cite{golse2005boltzmann, bardos1991fluid} of $\mathcal L$. Instead, guided by linearity and rotational invariance, we expand $\Phi^{(1)}$ in irreducible tensorial representations of $O(2)$, allowing for both even and odd two-dimensional tensors~\cite{fruchart2022odd}:
\begin{equation}
    \begin{split}
    \Phi^{(1)} = &\Big(\mathcal D(\bm c^2)\mathbf D+\mathcal D^\perp(\bm c^2)\mathbf D^\perp\Big):\bm \nabla\bm u~+\\
    &\Big(\mathcal A(\bm c^2)\mathbf A +\mathcal A^\perp(\bm c^2)\mathbf A^\perp\Big)\cdot\bm \nabla \log T\,,
    \end{split}
    \label{eq: definition of Phi}
\end{equation}
where we have introduced $\mathbf A^\perp = \bm\varepsilon \cdot \mathbf A$ and $\mathbf D^\perp = \bm\varepsilon \cdot \mathbf D$, together with the unknown scalar functions $\mathcal D$, $\mathcal D^{\perp}$, $\mathcal A$, and $\mathcal A^{\perp}$.
Although the left-hand side of Eq.~\eqref{eq: to solve} depends only on $\mathbf D$ and $\mathbf A$, chirality permits the operator $\mathcal L$ to mix these tensors with their transverse counterparts, coupling $\mathbf D$ to $\mathbf D^{\perp}$ and $\mathbf A$ to $\mathbf A^{\perp}$, in contrast to nonchiral systems~\cite{eren2025collisional}. This coupling underlies the emergence of nonzero odd transport coefficients. Since Eq.~\eqref{eq: to solve} contains no scalar coefficient on its left-hand side, $\Phi^{(1)}$ does not include any term proportional to $\bm\nabla \cdot \bm u$, consistent with the vanishing bulk viscosity of dilute smooth monatomic gases~\cite{Kremer_Santos_Garzó_2014,chapman1990mathematical}.

To proceed, we assume the unknown functions to be constants, namely $\mathcal D(\bm c^{2}) = \mathcal D_{0}$, $\mathcal D^\perp(\bm c^{2}) = \mathcal D_{0}^\perp$, $\mathcal A(\bm c^{2}) = \mathcal A_{0}$, and $\mathcal A^\perp(\bm c^{2}) = \mathcal A_{0}^\perp$. In principle, these functions could be expanded in a polynomial basis to retain their velocity dependence; however, higher-order terms are typically negligible~\cite{chapman1990mathematical}.
By linearity and rotational invariance, $\mathcal L$ does not couple irreducible tensors of different rank, so each sector can be treated independently. We therefore first focus on the shear sector associated with $\mathbf D$ and $\mathbf D^\perp$.

\subsection{Order \texorpdfstring{$\mu^1$}{1}: obtaining \texorpdfstring{$\mathcal D_0$}{D0} and \texorpdfstring{$\mathcal D_0^\perp$}{D0perp}}

To determine $\mathcal D_0$ and $\mathcal D_0^\perp$, we introduce the matrix elements of the linearized collision operator
\begin{equation}
    \mathsf L_{\alpha \beta}
    \equiv
    \Big\langle \big(f^{(0)}\big)^{-1}D^{(\alpha)}_{ij}, \mathcal L[D^{(\beta)}_{ij}] \Big\rangle\,,
    \quad
    \alpha,\beta\in\{\parallel,\perp\}\,,
\end{equation}
with scalar product
\begin{equation}
    \langle \psi,\phi\rangle \equiv \int f^{(0)}(\bm c)\psi(\bm c)\phi(\bm c) d\bm c\,,
\end{equation}
and $D^{(\parallel)}_{ij}\equiv D_{ij}$, $D^{(\perp)}_{ij}\equiv D^\perp_{ij}$.
Factoring out the velocity-gradient tensor, the shear part of Eq.~\eqref{eq: to solve} yields
\begin{equation}
    \frac{f^{(0)}}{T}\mathbf D(\bm c)=\mathcal D_0\mathcal L\left[ \mathbf D(\bm c)\right]+\mathcal D_0^\perp\mathcal L\left[ \mathbf D^\perp(\bm c)\right]\,.
\end{equation}
By projecting onto the basis $\{D_{ij},D^\perp_{ij}\}$, we obtain
\begin{equation}
{\renewcommand{\arraystretch}{1.1}
\dfrac{1}{T}\begin{pmatrix}
   \langle D_{ij}, D_{ij}\rangle\\
   \langle D_{ij}, D_{ij}^\perp\rangle
\end{pmatrix}
=
\begin{pmatrix}
    \mathsf L_{\parallel\parallel} & \mathsf L_{\parallel\perp} \\
    \mathsf L_{\perp\parallel}     & \mathsf L_{\perp\perp}
\end{pmatrix}
\begin{pmatrix}
    \mathcal D_0 \\
    \mathcal D_0^\perp
\end{pmatrix}\,.
}
\end{equation}
Rotational invariance implies
$\mathsf L_{\perp\parallel}=-\mathsf L_{\parallel\perp}$ and $\mathsf L_{\perp\perp}=\mathsf L_{\parallel\parallel}$. In a parity-invariant system, $\mathcal P^{-1}\mathcal L \mathcal P=\mathcal L$ (with $\mathcal P$ the reflection operator) forbids mixing between $\mathbf D$ and $\mathbf D^\perp$, hence $L_{\parallel\perp}=0$ and parity-odd transport is excluded. Parity-odd transport therefore requires parity breaking~\cite{fruchart2022odd} and hence a non-self-adjoint $\mathcal L$. However, $\mathcal L$ can be non-self-adjoint for many reasons unrelated to parity breaking, and non-self-adjointness alone does not guarantee odd transport: generic non-self-adjoint collision operators---such as those in equilibrium dense gases~\cite{Resibois_1970} or in dissipative dilute granular gases~\cite{Soto_Risso_Brito_2014} ($\Delta=0$)---do not, by themselves, generate parity-odd transport.

Using the relations $\langle D_{ij},D_{ij}\rangle=4nT^{2}$ and $\langle D^\perp_{ij},D_{ij}\rangle=0$, we find:
\begin{equation}
    \mathcal D_0=4nT\dfrac{\mathsf{L}_{\parallel\parallel}}{\mathsf{L}_{\parallel\parallel}^2+\mathsf{L}_{\parallel\perp}^2}\,,\quad \mathcal D_0^\perp=4nT\dfrac{\mathsf{L}_{\parallel\perp}}{\mathsf{L}_{\parallel\parallel}^2+\mathsf{L}_{\parallel\perp}^2}\,.
    \label{eq: expression for D0}
\end{equation}
The evaluation of $\bm{\mathsf L}$ is straightforward but lengthy (see Appendix~\ref{sec: viscosity} for details) and leads to the following result:
\begin{flalign}
        \mathsf L_{\parallel\parallel}=&- \chi\sigma n^2T^2 \sqrt{\dfrac{\pi T}{m}}\left(\dfrac{m\Delta^2}{T} + (1 + \alpha)(7 - 3\alpha)\right)\nonumber\\
        =&
        -4\sqrt{\pi}\chi\sigma n^{2}m^{2}|\Delta|^{5}\frac{(2 - \alpha)(1 + \alpha)}{(1-\alpha^{2})^{5/2}}\,,\\
        \mathsf L_{\parallel\perp}=&~2\pi \chi\sigma n^{2}T^{2}\Delta(1-\alpha)\nonumber\\
        =&~2\pi\chi\sigma n^{2}m^{2}\Delta^{5}\frac{1}{(1-\alpha)(1 + \alpha)^{2}}\,.
    \label{eq: expression for L}
\end{flalign}
In both cases, the second line is obtained by substituting the steady-state temperature [Eq.~\eqref{eq: temperature iso}]. We nevertheless keep $T$, $\Delta$, and $\alpha$ explicit in the intermediate expressions. Treating them as independent is unphysical but useful for building intuition and for consistency checks in various limits, as we now do.

At fixed $T$, the achiral limit $\Delta\to0$ restores parity, so the cross coefficient vanishes: $\mathsf L_{\parallel\perp}\to0$. The longitudinal coefficient $\mathsf L_{\parallel\parallel}$ also goes to zero, but only once $T$ is replaced by its steady-state value, since $T\to0$ as $\Delta\to0$. If instead one sets $\Delta=0$ while keeping $T$ and $\alpha$ fixed, the term $\mathsf L_{\parallel\parallel}$ remains finite, formally recovering the dissipative hard-disk result~\cite{brilliantov2010kinetic} and, as $\alpha\to1$, the equilibrium hard-disk limit~\cite{chapman1990mathematical}.

The limit $\alpha \to 1$ is singular because $T \to \infty$. Substituting the steady-state temperature therefore causes $\mathsf L$ to diverge. At fixed $T$, however, $\mathsf L_{\parallel\perp} \propto 1 - \alpha$, so $\mathsf L_{\parallel\perp}$ remains nonzero near $\alpha = 1$ only due to the divergence of $T$. Without this divergence, the coefficient would vanish even when parity is broken by $\Delta$.
This surprising result originates from the microscopic collision rules: chirality allows a single collision to rotate $D$ into $D^\perp$. Yet, in the limit $\alpha \to 1$, each such contribution is canceled by its mirror counterpart in the relevant integrals. Dissipation lifts this cancellation because $\bm v'_{12}\cdot\hat{\boldsymbol\sigma}_{12} \neq -\bm v_{12}\cdot\hat{\boldsymbol\sigma}_{12}$, thereby leaving a finite coupling $\mathsf L_{\parallel\perp} \propto \Delta (1 - \alpha)$.

With $\mathcal D_0$ and $\mathcal D_0^\perp$ in hand, we can now compute the viscosities.

\subsection{Order \texorpdfstring{$\mu^1$}{1}: the viscosities}

\begin{figure*}
    \centering
    \includegraphics[width=0.95\linewidth]{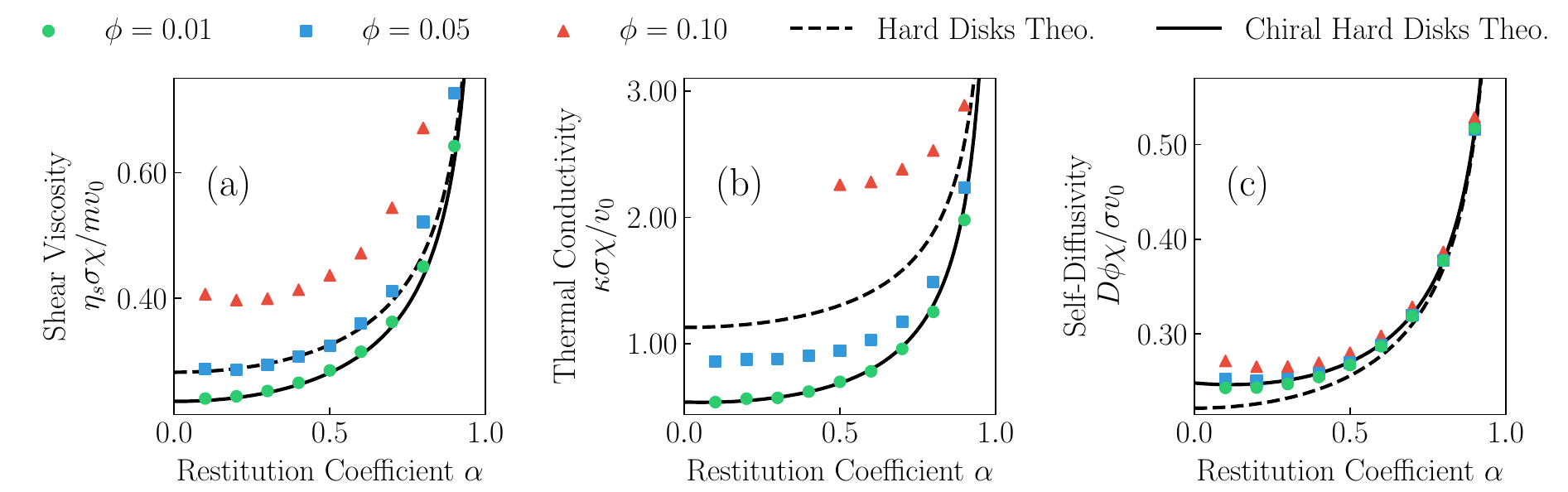}
    \caption{
    Even transport coefficients as functions of the restitution coefficient $\alpha$ at several densities $\phi$.
    (a) Shear viscosity measured from an imposed shear flow, as detailed in Appendix~\ref{app: viscosity measurement}.
    (b) Thermal conductivity measured by imposing a temperature gradient via a spatially dependent driving $\Delta = \Delta(x)$, as discussed in Appendix~\ref{app: conductivity measurement}. The missing data points for $\phi = 0.1$ correspond to systems in which the smallest numerically accessible thermal gradient was still too large to yield reliable measurements.
    (c) Self-diffusivity measured using a Green–Kubo relation (see Appendix~\ref{sec: diffusivity measurement} for details).
    Solid lines show the predictions of the chiral theory [Eqs.~\eqref{eq: viscosity theory both}, \eqref{eq: thermal conductivity all}, and \eqref{eq: diffusivity theo all}], while dashed lines indicate the expected values for regular hard disks at equilibrium temperature $k_B T^{\rm eq} = m \Delta^2/(1 - \alpha^2)$. The parameter $v_0$ denotes an arbitrary reference velocity, kept fixed across simulations to allow a direct comparison between the chiral and equilibrium cases without rescaling by the temperature, which provides the only natural velocity scale at equilibrium but varies with $\alpha$ in our system.
    }
    \label{fig: transport even}
\end{figure*}

With Eq.~\eqref{eq: kinetic def of stress tensor} and $f=f^{(0)}(1+\mu \Phi^{(1)})$, the stress splits into a zeroth-order (homogeneous) part and a first-order (viscous) correction:
\begin{equation}
    \begin{split}
        \bm \Pi&=-m\int \bm c\otimes\bm c f^{(0)}(1 + \mu\Phi^{(1)})d\bm c\\
               &=-nT\bm 1-\mu m\int \bm c\otimes\bm c f^{(0)}\Phi^{(1)} d\bm c\,.
    \end{split}
    \label{eq: stress to get visco}
\end{equation}
Substituting $\Phi^{(1)}$ [Eq.~\eqref{eq: definition of Phi}] into Eq.~\eqref{eq: stress to get visco}, we obtain
\begin{equation}
    \begin{split}
    \bm \Pi_{ij}^{\rm viscous}=&-\left(m\mathcal D_0\int f^{(0)}c_i c_j D_{kl}d\bm c\right)\partial_k u_l\\
    &-\left(m\mathcal D_0^\perp\int f^{(0)} c_i c_j D_{kl}^\perp d\bm c\right)\partial_k u_l\\
    =&-nT^2\mathcal D_0(\delta_{ik}\delta_{j l}+\delta_{il}\delta_{jk}-\delta_{ij}\delta_{kl})\partial_k u_l\\
    &+nT^2\mathcal D_0^\perp(\varepsilon_{ik}\delta_{j l}+\varepsilon_{j l}\delta_{i k})\partial_k u_l\,,
    \end{split}
\end{equation}
which, upon identification with the viscosity tensor [Eq.~\eqref{eq: viscosity decomposition}], yields
\begin{equation}
    \eta_s = -nT^2\mathcal D_0, \quad \eta_o=nT^2\mathcal D_0^\perp,\quad \zeta=\eta_A=\eta_B=\eta_R=0.
    \label{eq: expression for all viscosities}
\end{equation}
We note that the primary effect of chirality is to generate an odd viscosity $\eta_o$. By contrast, the antisymmetric viscosities $\eta_A$ and $\eta_R$ vanish because the kinetic stress is symmetric, while $\zeta$ and $\eta_B$ are zero since $\Phi^{(1)}$ contains no scalar $\bm\nabla \cdot \bm u$ sectors. However, these coefficients are expected to acquire collisional contributions in the non-dilute limit.

Using Eqs.~\eqref{eq: expression for D0}, \eqref{eq: expression for L} and \eqref{eq: expression for all viscosities}, we finally obtain
\begin{subequations}
\begin{flalign}
\label{eq:eta_s_expr}
\eta_s&=\frac{4m|\Delta| (2-\alpha)}
{\chi\sigma\sqrt{\pi}\sqrt{1-\alpha^{2}}P(\alpha)}\\
\eta_o&=\frac{2 m\Delta (1-\alpha)}
{\chi\sigma(1 + \alpha)P(\alpha)}\,,
\end{flalign}
\label{eq: viscosity theory both}
\end{subequations}
with $P(\alpha)$ a positive polynomial in $0\leq \alpha \leq 1$:
\begin{equation}
    P(\alpha)= 4(2-\alpha)^2(1+\alpha) + \pi(1-\alpha)^3\,.
\end{equation}
We note that the viscosity $\eta_s$ can be rewritten as
\begin{equation}
\eta_s = \frac{4 (2-\alpha)}{ \chi \sigma P(\alpha)} \sqrt{\frac{mT_\text{Gauss}}{\pi}} \,,
\end{equation}
upon substituting $T_\text{Gauss}$ (Eq.~\eqref{eq: temperature iso}) into Eq.~\eqref{eq:eta_s_expr}.
This compact expression shows that chirality (through $\Delta$) primarily renormalizes the conventional viscosity $\eta_s$ via the temperature $T_\text{Gauss}$. Moreover, in the elastic limit $\alpha \to 1$ ($P(\alpha \to 1) = 8$), it reproduces the standard equilibrium result for the viscosity of a hard-disk gas~\cite{chapman1990mathematical}.

Figure~\ref{fig: transport even}(a) compares simulations with Eq.~\eqref{eq: viscosity theory both} for the shear viscosity, showing excellent agreement at low $\phi$. Theory and simulations for $\eta_o$ were already compared in Fig.~\ref{fig: transport odd}(a); as noted there, $\eta_o \to 0$ as $\alpha \to 1$. This indicates that, in this model, odd viscosity requires dissipative normal collisions. Indeed, in the limit $\alpha \to 1$, the temperature diverges, so that $m\Delta^{2}/T \sim 1 - \alpha^{2} \to 0$. As a result, chirality becomes asymptotically negligible and $\mathsf L_{\parallel\perp}/\mathsf L_{\parallel\parallel} \to 0$ [Eqs.~\eqref{eq: expression for L}], causing $\eta_o$ to vanish.
An alternative argument can be made at fixed $T$: in that case, $\mathsf L_{\parallel\perp} \propto T^2\Delta(1-\alpha)$, which directly implies $\eta_o \propto 1 - \alpha$ [Eq.~\eqref{eq: odd visco vanish before}].
However, with this model, the divergence of $T$ as $\alpha\to1$ masks this effect, and the vanishing of the odd viscosity has to be found in the previous argument. This mechanism becomes relevant when $T$ remains finite as $\alpha\to 1$, for example, because of external damping. We return to this point in the conclusion.

\subsection{Order \texorpdfstring{$\mu^1$}{1}: the thermal conductivities}

Applying the same Chapman-Enskog procedure in the thermal sector yields the regular and odd thermal conductivities, $\kappa$ and $\kappa_o$. As detailed in Appendix~\ref{sec: conductivity}:
\begin{subequations}
\begin{flalign}
    &\kappa=\frac{16|\Delta|(1 + \alpha)(4-3\alpha)}{\chi \sigma \sqrt{\pi}\sqrt{1-\alpha^{2}}Q(\alpha)}\\
    &\kappa_o= \frac{8 \Delta \left(1-\alpha \right)}{\chi \sigma Q(\alpha)}\,,
\end{flalign}
\label{eq: thermal conductivity all}
\end{subequations}
where $Q$ is a positive polynomial given in Appendix~\ref{sec: conductivity}. As for viscosity, chirality generates the odd conductivity $\kappa_o$ and affects the conventional conductivity $\kappa$.
In the elastic limit $\alpha\to1$, $\kappa$ diverges due to the divergence of the steady-state temperature, while $\kappa_o\to0$. 
Figs.~\ref{fig: transport even}(b) and \ref{fig: transport odd}(b) show good agreement with simulations at low $\phi$. The small discrepancies between the theory and simulations for the odd thermal conductivity are likely attributable to the difficulties in measuring this transport coefficient, as explained in Appendix~\ref{app: conductivity measurement}.

By substituting the temperature $T_\text{Gauss}$ (Eq.~\eqref{eq: temperature iso}) into Eq.~\eqref{eq:kappa_expr}, we obtain a compact expression for the thermal conductivity:
\begin{equation}
\label{eq:kappa_expr_eq}
\kappa = \frac{16 (1+\alpha)(4-3\alpha)}{ \chi \sigma Q(\alpha)} \sqrt{\frac{T_\text{Gauss}}{\pi m}} \,.
\end{equation}
In the elastic limit $\alpha\to 1$ ($Q(\alpha \to 1) = 16$), this expression reduces to the results for a dilute hard-disk gas~\cite{chapman1990mathematical}. It shows that chirality (through $\Delta$) primarily affects the thermal conductivity $\kappa$ by renormalizing it via the temperature $T_\text{Gauss}$.

\subsection{Self-diffusivities via Chapman-Enskog}

The self-diffusivity $D$ and its odd counterpart $D_o$ follow from a Chapman-Enskog expansion of the Boltzmann-Lorentz equation for a tagged particle~\cite{brilliantov2010kinetic}. In this framework, the nonlinear bilinear operator ${\mathcal J}(\mathbf r,\bm v|f,f)$ is replaced by the linear tracer-bath operator ${\mathcal J}(\mathbf r,\bm v| f_s,f_B)$, where only the tracer distribution $f_s$ evolves and ‘‘collides'' with the bath distribution $f_B\equiv f^{(0)}(\bm v)$.

Performing a Chapman–Enskog expansion on $f_s$, we obtain the self-diffusivities~\cite{gomez2024diffusion} (see Appendix~\ref{sec: diffusivity} for the derivation):
\begin{subequations}
\begin{flalign}
\label{eq:D_expr_eqlike}
    &D=\dfrac{\sqrt{\pi}\sigma|\Delta|}{\phi\chi\sqrt{1-\alpha^2}R(\alpha)}\\
    &D_o=\dfrac{\pi\sigma\Delta}{2\phi\chi(1 + \alpha)R(\alpha)}\,,
\end{flalign}
\label{eq: diffusivity theo all}
\end{subequations}
with $R(\alpha)$ a strictly positive polynomial given in Appendix~\ref{sec: diffusivity}.

As in the case of viscosity ($\eta_s$) and thermal conductivity ($\kappa$), chirality introduces off-diagonal antisymmetric elements in the diffusion matrix (odd diffusivity $D_o$) and renormalizes the diagonal self-diffusivity $D$ through the temperature $T_\text{Gauss}$ (Eq.~\eqref{eq: temperature iso}).
By substituting Eq.~\eqref{eq: temperature iso} into Eq.~\eqref{eq:D_expr_eqlike}, the self-diffusivity can be expressed as
\begin{equation}
D = \frac{4}{n \chi \sigma R(\alpha)} \sqrt{\frac{T_\text{Gauss}}{\pi m}} \,.
\end{equation}
As expected, this compact expression shows that $D$ reduces to the standard hard-disk result~\cite{brilliantov2010kinetic} in the elastic limit $\alpha \to 1$, for which $R(\alpha \to 1) = 8$.

As a consequence of the temperature divergence in the elastic limit, the self-diffusion $D$ diverges, while $D_o$ remains finite as $\alpha\to1$. 
Indeed, each chiral collision imparts an oriented transverse kick to the tagged particle, so the mean transverse response---and the corresponding integrals---remain finite even as $\alpha \to 1$. However, as for odd viscosity and odd conductivity, the ratio of off-diagonal to diagonal matrix elements of the linearized Boltzmann–Lorentz operator still vanishes in this limit. 
%For diffusion, this vanishing is weaker than for the other coefficients: the odd diffusion coefficient $D_o$ remains finite, while the conventional diffusion coefficient $D$ diverges.
The comparison between simulations and theory shows excellent agreement across a broad range of densities within the low-density regime, as illustrated in Fig.~\ref{fig: transport even}(c) and Fig.~\ref{fig: transport odd}(c). At higher packing fractions, however, both the odd and even self-diffusion coefficients tend to increase slightly.

Although the self-diffusivity is well defined, the system as a whole does not diffuse in the absence of an external bath. A natural extension would therefore be to add noise and study how the collective diffusivity depends on the packing fraction. In the strongly chiral regime, this dependence was recently found to be non-monotonic~\cite{guo2025diffusion}, as for the self-diffusion of active chiral tracers in passive baths~\cite{luigi2025self}.

\section{Discussion and outlook}

Because our model is based on hard disks, the interaction is singular. Nevertheless, we expect the qualitative phenomenology derived here to persist for chiral fluids with short-ranged forces. In particular, systems with softer repulsive cores and short-ranged chiral interactions should display comparable behavior~\cite{caprini2025Bubble}. Indeed, one may start from a smooth $N$-particle dynamics, write a Liouville equation for the full phase-space distribution, and derive a Boltzmann description by truncating the associated BBGKY hierarchy~\cite{Marconi2026hydrodynamics}.

It is also straightforward to incorporate damping $\gamma$, possibly together with a Langevin bath at temperature $T_b$. Between collisions, the velocities would obey $\dot {\bm v}=-\gamma \bm v+\sqrt{{2\gamma T_b}/{m}}\bm\zeta(t)$,
so that the Boltzmann equation acquires the additional contribution $\gamma\partial_{\bm v}\cdot(\bm v f)+{\gamma T_b}/{m}\partial_{\bm v}^2 f$. Within the Chapman-Enskog expansion for a Gaussian homogeneous solution, this modification amounts to a simple shift of the linearized operator $\mathcal L[\bullet]\to \mathcal L[\bullet]+\gamma\partial_{\bm v}\cdot(\bm v\bullet)+{\gamma T_b}/{m}\partial_{\bm v}^2(\bullet)$. These new terms mainly change the steady-state temperature and shift the diagonal matrix elements of $\mathcal L$ after integration by parts~\cite{garzo2002transport}. In the shear sector, for instance, $\mathsf L_{\parallel\parallel}\to \mathsf L_{\parallel\parallel}-8\gamma nT^2$ while $\mathsf L_{\parallel\perp}$ is unchanged and the viscosities still follow from Eq.~\eqref{eq: expression for all viscosities}. Here $T$ is determined from Eq.~\eqref{eq: temperature} as a stable positive root of the cubic equation~\cite{maire2024non}:
$0=\omega(\phi,T)\langle \delta E\rangle_{\rm coll}/2-2\gamma (T-T_b)$.

A limitation of our current computation is the dilute-gas approximation. It neglects collisional-transfer contributions to transport and, crucially, cannot capture antisymmetric stresses generated by collisions at finite densities. Yet this is precisely the experimentally relevant regime for most chiral systems such as crowded biological media~\cite{shi2025ordering, chen2024emergent, tan2022odd, li2024robust}. Extending the theory beyond the dilute limit is thus essential to bridge this gap and identify the mechanisms that control dense chiral transport. Notably, it would lead to the first analytical expressions for the other odd transport coefficients, such as $\eta_A$ or $\eta_R$.

The dense regime is also where correlations and fluctuations become unavoidable, motivating a fluctuating kinetic description beyond the Boltzmann-Grad limit: a fluctuating Boltzmann equation~\cite{dorfman2021contemporary, bouchet2020boltzmann, bodineau2023statistical}. Non-Gaussian corrections to the velocity distribution captured by the Sonine expansion are known to violate the fluctuation–dissipation relation for the associated noise in the fluctuating Boltzmann equation and to induce memory effects in the hydrodynamic noise~\cite{brey2009fluctuating, maynar2009fluctuating}. We expect these corrections to strongly influence the fluctuating hydrodynamics of chiral particles~\cite{han2021fluctuating}.

Beyond fluids, in dense regimes, chirality can also produce an antisymmetric elastic response known as odd elasticity in amorphous or crystalline materials~\cite{scheibner2020odd}. In equilibrium, hard-disk systems can be systematically coarse-grained to derive a hydrodynamic description of a crystal~\cite{kirkpatrick1990kinetic}. Carrying out an analogous program for active chiral crystals would be particularly timely, as these systems display a rich set of instabilities that depend sensitively on transport coefficients and elastic moduli~\cite{lee2025odd}.

More broadly, while we focused on the self-diffusivity of a tagged particle identical to the bath constituents, an immediate extension is to consider a geometrically complex tracer immersed in a bath of chiral colliders~\cite{cleuren2007granular,costantini2007granular,hargus2025odd,Passive2025HargusPRE, poggioli2023odd}. By tuning the tracer shape, one can implement symmetry breaking in a controlled way and generate a hierarchy of nonequilibrium transport effects, from granular-ratchet drift~\cite{costantini2007granular,cleuren2007granular} to autonomous rotation~\cite{hargus2025odd,Passive2025HargusPRE}. 

Finally, extending these calculations to 3D is natural, but the symmetry constraints are qualitatively different from 2D~\cite{chatzittofi2025dumbbell, markovich2021odd}. In a fully isotropic 3D fluid, an odd viscosity is forbidden---no fourth-rank isotropic tensor can be built from a single $\varepsilon_{ijk}$---and the torque density $\tau_k=\varepsilon_{kij}\Pi_{ij}/2$ is a pseudovector that must vanish by isotropy. Odd viscosity and a nonzero $\bm\tau$ become possible only if isotropy is broken by a distinguished axis $\hat{\bm\ell}$ (e.g., selecting the plane of transverse kicks $\Delta$), in which case one may write $\eta^{\rm odd}_{ijkl}=\eta_o\big(\varepsilon_{ikm}\hat{\ell}_m\delta_{jl}+\varepsilon_{jkm}\hat{\ell}_m\delta_{il}\big)$ and $\bm\tau=\tau\hat{\bm\ell}$. In principle, the additional contractions enabled by $\hat{\bm \ell}$ even allow for a richer set of transport coefficients than in 2D.

\subsection{Acknowledgments}

L.C. acknowledges financial support from the University of Rome Sapienza under the project Ateneo 2024 (Grant No. RM124190C54BE48D), ‘‘Elementary excitations at the origin of glassy or hexatic behavior in low dimensional system, at and out of equilibrium.''

\appendix

\section{Numerical measurement of transport coefficients}\label{sec: transport}
\subsection{Stress and heat current in event-driven molecular dynamics}
The stress and heat current are generically given by a kinetic part plus an impulsive collisional contribution. For hard-disks with instantaneous collisions, their microscopic expressions read~\cite{irving1950statistical}
\begin{equation}
    \begin{split}
        \bm\Pi(t)=&-\frac{m}{L^{2}}\sum_{\alpha=1}^{N}\bm c_{\alpha}(t)\otimes \bm c_{\alpha}(t)
        \\
        &+\frac{\sigma}{L^{2}}\sum_{\alpha<\beta}\hat{\bm\sigma}_{\alpha\beta}(t)\otimes \mathbf I_{\alpha\beta}(t)
        \delta\left(t-t^{\rm coll}_{\alpha\beta}\right)\,,\\
        \bm J(t)=&+
        \frac{m}{2L^{2}}\sum_{\alpha=1}^{N} \bm c_{\alpha}^{2}(t)\bm c_{\alpha}(t)\\
        &
        -\frac{\sigma}{L^{2}}\sum_{\alpha<\beta}\Big[\mathbf I_{\alpha\beta}(t)\cdot\bm C_{\alpha\beta}(t)\Big]
        \hat{\bm\sigma}_{\alpha\beta}(t)
        \delta\left(t-t^{\rm coll}_{\alpha\beta}\right)\,,
        \end{split}
        \label{eq: stress and heat}
\end{equation}
where $\bm c_{\alpha}=\bm v_{\alpha}-\langle\bm v\rangle$, $\bm C_{\alpha\beta}=(\bm c_{\alpha}+\bm c_\beta)/2$ and $\mathbf I_{\alpha\beta}$ is the impulse transferred during the collision between $\alpha\beta$,
$\mathbf I_{\alpha\beta}\equiv m\big(\bm v'_{\alpha}-\bm v_{\alpha}\big)=-m\big(\bm v'_{\beta}-\bm v_{\beta}\big)$.

Transport coefficients are often expressed through Green–Kubo formulas with time integrals of current autocorrelations. However, the instantaneous microscopic currents contain impulsive collisional contributions, which would never be sampled via instantaneous measurement in the simulations, and only the kinetic term would be retained. Moreover, in driven, nonequilibrium steady states Green–Kubo relations are not guaranteed to hold in their equilibrium form~\cite{dufty2002green}. We therefore avoid this route whenever an alternative, more direct evaluation is available.

\subsection{Viscosities}
\label{app: viscosity measurement}

The viscosities are measured under a controlled simple shear of rate $\dot\gamma$ imposed via Lees-Edwards boundary conditions~\cite{lees1972computer} in a box of size $L_x\times L_y$ where the periodic images above and below the simulation cell slide along $\hat{x}$ with velocities $\pm \dot\gamma L_y/2$ (equivalently, a relative velocity $\dot\gamma L_y$ between the top and bottom images). This introduces a time-dependent shear offset $\delta x(t)=\dot\gamma tL_y$ between adjacent replicas. Accordingly, when a particle crosses the top or bottom boundary, its position is remapped as $y\to y\mp L_y$ together with an affine shift $x\to x\mp \delta x(t)~ (\mathrm{mod}~ L_x)$. 

Following Eqs.~\eqref{eq: hydro intro}, such boundary conditions impose a shear profile:
\begin{equation}
\bm{u}(y)=\dot\gamma y\hat{\bm{x}}\,,
\end{equation}
which implies:
\begin{equation}
    \begin{gathered}
    \partial_y u_x = \dot\gamma,\\
    \partial_x u_x = \partial_x u_y = \partial_y u_y = 0.
    \end{gathered}
\end{equation}
Using Eq.~\eqref{eq: viscosity from pressure}, we find:
\begin{equation}
    \eta_s=\frac{\langle \Pi_{xy} + \Pi_{yx}\rangle}{2\dot\gamma},\qquad\eta_o=\frac{\langle \Pi_{xx}-\Pi_{yy}\rangle}{2\dot\gamma}\,.
\end{equation}
Since $\langle \Pi \rangle$ can be computed using a time average, the singular contributions of the collisions to the \textit{instantaneous} stress are smoothed out to simple jumps in values~\cite{mabillard2024hydrodynamic}.

A drawback of this approach is that the range of $\dot\gamma$ over which the response is truly linear is not known a priori. In practice, we had to use an extremely low shear rate, $\sigma\dot\gamma/\Delta=2\times10^{-6}$, because at larger values we observed a pronounced $\dot\gamma$ dependence of $\eta_o$. Beyond genuine nonlinearities, this sensitivity may also reflect the fact that even a nonchiral fluid under shear develops a nonzero normal-stress difference $\Pi_{xx}-\Pi_{yy}$. This term is typically neglected since in linear theories it appears only at Burnett order~\cite{saha2016normal}, $\mathcal{O}(\bm\nabla^3)$, but the small magnitude of $\eta_o$ (despite entering already at Navier-Stokes order) likely makes it comparable to this Burnett-level contribution unless $\dot\gamma$ is taken extremely small. We note that $\eta_s$ could be accurately measured with $\sigma\dot\gamma/\Delta \sim \mathcal O(10^{-3})$ for all our densities.

The viscosities are measured in systems of $N=1000$ particles for a time $1.3\times 10^9\leq t\Delta/\sigma\leq 5\times 10^{9}$ with between 5 and 50 independent realizations.

\subsection{Conductivities}
\label{app: conductivity measurement}
It is possible to measure the thermal conductivities by imposing a thermal gradient. However, a gradient generated by fixing $T$ at two slabs (e.g., via a M\"uller-Plathe scheme~\cite{muller1997simple}) is screened by bulk relaxation due to energy non-conservation during collision. Therefore, it is easier to impose the thermal gradient by directly fixing the temperature $T(x)\simeq m\Delta^2(x)/(1 - \alpha^2)$ at each point by varying $\Delta$ with $x$. For example, we can ask that the temperature profile has a triangular shape: $T(x)=T_0+{\delta T}/{2}-{2\delta T}\left|x-{L_x}/{2}\right|/{L_x}$ which linearly interpolates between $T(0)=T(L_x)=T_0 - \delta T/2$ and $T(L_x/2)=T_0+\delta T/2$. Such a profile is imposed by using a collision rule dependent on the position of the two particles 1 and 2 with $\Delta_{12} = \sqrt{T(\,(x_1+x_2)/2\,)(1 - \alpha^2)/m}$.  Assuming $\delta T$ to be small enough, $\partial_x T|_{x \simeq L_x/4}=-\partial_x T|_{x \simeq 3L_x/4}=\rm{cst}$ and the conductivities can be obtained by measuring the local heat current around these points. This method works well for the regular thermal conductivity, but not for its odd counterpart. Indeed, the largest $\delta T$ we can use while maintaining an acceptable signal-to-noise ratio still does not yield a converged value, as a slight increase in $\delta T$ changes the extracted odd conductivity. One possible reason is that, under periodic boundary conditions, imposing a temperature gradient while enforcing $u_x=0$ can generate a density gradient through the requirement of mechanical equilibrium (uniform pressure). This may contaminate the measurement because the resulting torque density may not remain uniform and can therefore induce a transverse flow ($u_y\neq 0$). In addition, if we retained Sonine corrections to the local stationary distribution so that $f^{(0)}$ were not strictly Gaussian, the heat current would acquire an additional contribution proportional to the density gradient~\cite{garzo2018enskog}:
\begin{equation}
    \bm{J}^{\rm non\text{-}Gaussian} = -\lambda\bm\nabla n - \lambda_o\bm\varepsilon\cdot\bm\nabla n\,,
\end{equation}
which would further interfere with our determination of the thermal conductivity. For these reasons, we adopt an alternative method to measure this coefficient.

As previously described, the measurement of the instantaneous heat current is impossible for discontinuous potentials due to the singular terms arising in its expression. The typical method to circumvent this problem involves the use of Helfand moments~\cite{helfand1960transport} related to the time integral of $J$, which smooths out the singular contributions. However, no Helfand relation is available for the odd thermal conductivity, and we have to rely on the standard Green-Kubo expression~\cite{hargus2025flux},
\begin{equation}
\kappa_o=\frac{L^{2}}{2T^{2}}\int_{0}^{\infty}
\Big\langle J_x(t)J_y(0)-J_y(t)J_x(0)\Big\rangle dt \,.
\label{eq: GK kappao}
\end{equation}
In practice, since $\bm J(t)$ contains impulsive contributions, we evaluate the correlator using the time-averaged current over a finite small bin $\Delta t$, $\bar{\bm J}(t)=\Delta t^{-1}\int_{t}^{t+\Delta t}\bm J(t')dt'$, and compute Eq.~\eqref{eq: GK kappao} with $\bm J\to\bar{\bm J}$. This binning is less problematic for $\kappa_o$ than for $\kappa$ (if we were to use a standard Green-Kubo for the latter), since the singular term~\cite{dufty2002shear} $\langle J_i(t)J_j(0)\rangle\propto\delta(0)$ cancels under antisymmetrization.

We see systematic discrepancies between the measurements and the theory for the thermal conductivities. Such discrepancies also appeared when we tried to evaluate the viscosities using Green-Kubo relations (with or without using Helfand moments). We also verified that no such strong discrepancies appeared in pure equilibrium hard disks for the regular thermal conductivity. This suggests that the issue is not the treatment of impulsive currents per se, but rather the limited applicability of equilibrium Green-Kubo relations in this nonequilibrium steady state~\cite{dufty2002green, brey2004simulation}. We also note that using $n\langle |T(\bm q\to 0)|^2\rangle$ instead of $T^2$ (with $T(\bm q)$ the Fourier transform of $T(\mathbf r)$) in the denominator, as suggested by Ref.~\onlinecite{hargus2025flux}, does not improve the result.

The regular thermal conductivity is measured in systems with $\delta T/m\Delta^2 = 2\times 10^{-3}$, $N = 10000$ and $L_x = 3L_y$ for a time $2.5\times 10^8\leq t\Delta/\sigma\leq 4\times 10^{8}$. 

The odd thermal conductivity is measured in systems of $N = 2000$ particles where the autocorrelations are computed each $dt \Delta/\sigma=1$ over a window of $t\Delta/\sigma = 2\times 10^6$ and averaged over 10 independent realizations.

\subsection{Self-diffusivities}
\label{sec: diffusivity measurement}
The ordinary and odd self-diffusivities are obtained from the velocity-displacement correlations introduced in Ref.~\onlinecite{hargus2021odd}:
\begin{equation}
    \begin{split}
        D&=\lim_{t\to\infty}\dfrac{\Big\langle \big( x(t) - x(0)\big)v_x(0)+\big( y(t) - y(0)\big)v_y(0)\Big\rangle}{2}\,,\\
        D_o&=\lim_{t\to\infty}\dfrac{\Big\langle \big( x(t) - x(0)\big)v_y(0)-\big( y(t) - y(0)\big)v_x(0)\Big\rangle}{2}\,.
    \end{split}
\end{equation}
Unlike stress or heat current autocorrelations in hard-particle systems, these estimators are free of impulsive singularities and can be evaluated directly without special regularization. %Compared to the Green-Kubo formulas for the viscosities and conductivity, these Green–Kubo formulas are exact even in the nonequilibrium setting. 

The diffusivities are measured in systems of $N = 200$ particles where the autocorrelations are computed each $dt \Delta/\sigma=10$ over a window of $t\Delta/\sigma = 0.5\times 10^9$ and averaged over 10 independent realizations.

\section{Sonine expansion for the homogeneous solution}\label{sec: sonine}
The homogeneous distribution $f(\bm v)$ must be obtained from the stationary Boltzmann equation:
\begin{equation}
    {\mathcal J}(\bm v| f, f) = 0\,.
    \label{eq: solution to f appendix}
\end{equation}
Using Eq.~\eqref{eq: J dense}, the stationarity condition ${\mathcal J}(\bm v|f,f)=0$ follows from a global balance between loss and gain terms. It is easier to check for the more restrictive pointwise (effective) detailed balance: 
\begin{equation}
    f( \bm v_1'')f(\bm v_2'')=\alpha^2f(\bm v_1)f(\bm v_2).
\end{equation}
At equilibrium $\alpha\to 1$ and $\Delta \to 0$, such a relation is \textit{exactly} satisfied by a Gaussian. We can try to see how far away we are from the Gaussian in our case:
\begin{equation}
    \begin{split}
        \alpha^{2}\dfrac{f^G( \bm v_1)f^G(\bm v_2)}{f^G(\bm v_1'')f^G(\bm v_2'')}&=\alpha^{2}\exp\Bigg(\dfrac{\bm v_1''^2 + \bm v_2''^2-\bm v_1^2 - \bm v_2^2 }{2T/m}\Bigg)\\
        &=\alpha^{2}\exp\Bigg(\dfrac{(\alpha^{-2} - 1)(\bm v_{12} \cdot \hat{\bm \sigma}_{12})^2 }{2T/m}\\
        &\qquad\qquad\quad+\dfrac{2\Delta^2 +2\Delta\bm v_{12}\cdot \hat{\bm \sigma}_{12}^\perp}{2T/m}\Bigg)\,.
    \end{split}
\end{equation}
The right-hand side must approach 1 if a Gaussian is a solution. In the limit $\alpha\to 1$ and $\Delta \to 0$, since collisions conserve energy, the Gaussian is exactly a solution. In the case $\alpha\neq 1$ and $\Delta \neq 0$, a Gaussian typically does not satisfy Eq.~\eqref{eq: solution to f appendix}. However, we saw that $T = m\Delta^2/(1 - \alpha^2)$ and since $\bm v_{12}\sim \mathcal O(\sqrt{T})$, the two terms proportional to $\Delta$ and $\Delta^2$ become negligible in the limit $\alpha\to 1$ since $T\propto(1 - \alpha^2)^{-1}\to \infty$. Similarly, the term proportional to $\alpha^{-2}-1$ vanishes in this limit, leaving 
\begin{equation}
    \lim_{\alpha \to 1}\alpha^{-2}\dfrac{f^G( \bm v_1'')f^G(\bm v_2'')}{f^G(\bm v_1)f^G(\bm v_2)}=1\,.
\end{equation}
Therefore,  a Gaussian is a solution in the limit $\alpha\to 1$ because the energy increments of $\Delta$ at collision are negligible with respect to the system temperature $T$.

We now compute perturbative corrections to $f^{ G}$ around $\alpha\to 1$ using a Sonine expansion~\cite{brilliantov2010kinetic}:
\begin{equation}
    f(\bm v)=f^{G}(\bm v)\left(1 + a_1S_1(\bm v^2) + a_2S_2(\bm v^2)+\dots\right)\,,
\end{equation}
where the Sonine polynomials $S_p$ are orthogonal to each other with respect to the Gaussian weight. In 2D, we notably have~\cite{brilliantov2010kinetic}:
\begin{equation}
    \begin{split}
        S_1(\bm v^2)&=1-\bm v^2/v_T^2\,,\\
        S_2(\bm v^2)&=1-2\bm v^2/v_T^2+\bm v^4/(2v_T^4)\,,
    \end{split}
\end{equation}
with $v_T = \sqrt{2T/m}$. The temperature definition $T \equiv \frac{m}{2n}\int \bm v^2f(\bm v)d\bm v$ imposes $a_1=0$ since $\int \bm v^2f(\bm v^2)d\bm v=2nT(1 - a_1)/m$, while the remaining non-trivial coefficients follow from the requirement that $f$ satisfy Eq.~\eqref{eq: solution to f appendix}.

Multiplying Eq.~\eqref{eq: solution to f appendix} by $\bm v^{2p}$ and integrating yields moment constraints,
\begin{equation}
    \begin{split}
        \int \bm v^{2p}{\mathcal J}(\bm v| f, f)d\bm v&=\dfrac{\sigma \chi}{2}\int  \Theta(-\boldsymbol v_{12}\cdot \hat{\boldsymbol \sigma}_{12})|\boldsymbol v_{12}\cdot \hat{\boldsymbol \sigma}_{12}|\\
        &\quad\times\Big((\bm v'_1)^{2p}+(\bm v'_2)^{2p}-\bm v_1^{2p}-\bm v_2^{2p}\Big)\\
        &\quad\times f(\boldsymbol v_1)f(\boldsymbol v_2) d\boldsymbol v_1d\boldsymbol v_2d\hat{\boldsymbol\sigma}_{12}\\
        &=0\,.
    \end{split}
\end{equation}
Evaluating these integrals with the method presented in Appendix.~\ref{sec: kinetic inte} (see also SM) and truncating at $a_2$ gives, for $p=1$,
\begin{equation}
    (1-\alpha^{2})T-m\Delta^{2}+a_{2}\dfrac{m\Delta^{2}+3(1-\alpha^{2})T}{16}+\mathcal{O}(a_2^2)=0\,,
\end{equation}
and, for $p=2$,
\begin{equation}
    \begin{split}
    &\Big(m\Delta^2\big[m\Delta^2 + T(2\alpha^2+7)\big]+ T^2(\alpha^2-1) (2\alpha^2+7)\Big)\\
    &-\dfrac{a_2}{16}\Big(T^2(1 + \alpha)\left(30\alpha^2(\alpha-1)+177\alpha - 209\right) -\\&\qquad m\Delta^2\big[m\Delta^2-T(6\alpha^2+29)\big]\Big)+\mathcal{O}(a_2^2)=0\,.
    \end{split}
\end{equation}
Solving these two equations determines $T(\alpha, m\Delta^2)$ and $a_2(\alpha, m\Delta^2)$ at this order, leading to the theoretical prediction given in Fig.~\ref{fig: iso}. The solution is density-independent because the sole density dependence enters through $\chi$, which reduces to a constant prefactor in $J$ for homogeneous states.

Other closures are possible (e.g., using $p=1$ and $p=3$ to get $a_2$ and $T$), but lower-order moments typically provide the most robust truncation.

\section{Chapman-Enskog methods}
\subsection{Derivation of the hydrodynamic equations}\label{sec:balance}

To derive the hydrodynamics equations for $n$, $\bm u$ and $T$ [Eqs.~\eqref{eq: hydro intro}], we start from the Boltzmann-Enskog equation~\eqref{eq: boltzmann equation}:
\begin{equation}
\partial_t f(\mathbf r,\bm v,t)+\bm v\cdot \bm\nabla f(\mathbf r,\bm v,t)={\mathcal J}(\mathbf r,\bm v|f,f)\,.
\end{equation}
For any test function $A(\bm v)$, multiplying by $A$ and integrating over $\bm v$ gives
\begin{equation}
\partial_t \int A f d\bm v+\bm\nabla\cdot\int A\bm v fd \bm v=\int A\mathcal Jd\bm v\,.
\label{eq: moment balance general}
\end{equation}
\textbf{Density}---With $A=1$, mass conservation by collisions implies $\int\mathcal Jd\bm v=0$ and~\eqref{eq: moment balance general} yields
\begin{equation}
\partial_t n+\bm\nabla\cdot(n\bm u)=0\,,
\end{equation}
corresponding to the continuity equation \eqref{eq:continuityEq}.

\vskip5pt
\noindent
\textbf{Momentum}---With $A=m\bm v$, Eq.~\eqref{eq: moment balance general} gives
\begin{equation}
\partial_t(mn\bm u)+\bm\nabla\cdot\left(m\int \bm v\otimes\bm v fd\bm v\right)=m\int \bm v\mathcal J d\bm v\,.
\end{equation}
Writing $\bm v=\bm u+\bm c$ and defining the kinetic stress $\bm\Pi^{\rm kin}=-m\int \bm c\otimes \bm c fd\bm v$ and its collisional counterpart $\bm \Pi^{\rm coll}$ via $\bm \nabla \cdot \bm \Pi^{\rm coll}=m\int \bm v \mathcal J d\bm v$, one obtains
\begin{equation}
\partial_t \bm u+\bm u\cdot\bm\nabla \bm u=\frac{1}{mn}\bm\nabla\cdot (\bm\Pi^{\rm kin} + \bm \Pi^{\rm coll})\,.
\end{equation}
The collisional stress stems from collisional momentum transfer over the contact displacement, even though each collision locally conserves $m\bm v$. In the dilute limit used in the main text, collisions are treated as occurring at the same point, so this contribution is neglected. We also note that $\bm \Pi^{\rm homo}$ and $\bm \Pi^{\rm visc}$ both get kinetic and collisional contributions.

\vskip5pt
\noindent
\textbf{Temperature}---With $A=m\bm c^2/2$, Eq.~\eqref{eq: moment balance general} yields
\begin{equation}
\partial_t T+\bm u\cdot\bm\nabla T=\frac{1}{n}\Big(\bm\Pi:\bm\nabla \bm u-\bm\nabla\cdot \big(\bm J^{\rm kin}+\bm J^{\rm coll}\big)\Big)+\delta\dot T\,,
\label{eq: app temp balance}
\end{equation}
with $\bm J^{\rm kin}\equiv m\int c^2 \bm c fd\bm v/2$ and $-\bm \nabla\cdot\bm J^{\rm coll}+n\delta \dot T=m\int \bm c^2\mathcal Jd\bm v/2$. In the homogeneous state $\bm J=0$ and $\bm u=0$, and
 Eq.~\eqref{eq: app temp balance} reproduces Eq.~\eqref{eq: temperature}.
\subsection{Kinetic integrals}\label{sec: kinetic inte}
We repeatedly encounter matrix elements of the linearized collision operator of the form
\begin{equation}
    \begin{split}
        \langle \phi \big(f^{(0)}\big)^{-1},\mathcal L[\psi]\rangle\equiv &\int \phi(\boldsymbol c_1)\mathcal L[\psi](\boldsymbol c_1)d\boldsymbol c_1\\
        =&\chi\sigma\int  \Theta(-\boldsymbol c_{12}\cdot \hat{\boldsymbol \sigma}_{12})|\boldsymbol c_{12}\cdot \hat{\boldsymbol \sigma}_{12}|\delta \phi\\&\times \psi(\boldsymbol c_2) f^{(0)}(\boldsymbol c_1)f^{(0)}(\boldsymbol c_2)  d\boldsymbol c_1d \boldsymbol c_2d\hat{\boldsymbol\sigma}_{12}\,,
    \end{split}
    \label{eq: kinetic integral}
\end{equation}
for any $\phi$ and $\psi$, and where:
\begin{equation}
    \delta \phi=\phi(\boldsymbol c_1')+\phi(\boldsymbol c_2') - \phi(\boldsymbol c_1)-\phi(\boldsymbol c_2)\,,
    \label{eq: change}
\end{equation}
is the change of any quantity $\phi(\bm c_1)+\phi(\bm c_2)$ at collision.

These integrals simplify in center-of-mass $\bm C = (\bm c_1 + \bm c_2)/2$ and relative $\bm c_{12}=\bm c_1-\bm c_2$. Using the collision rule~\eqref{eq: collision rule}:
\begin{equation}
    \begin{split}
        \boldsymbol c_1&=\boldsymbol C +\dfrac{1}{2}\boldsymbol c_{12}\,,\\
        \boldsymbol c_2&=\boldsymbol C -\dfrac{1}{2}\boldsymbol c_{12}\,,\\
        \boldsymbol c'_1&=\boldsymbol C +\dfrac{1}{2}\boldsymbol c_{12}-\dfrac{1+\alpha}{2}(\boldsymbol c_{12}\cdot\hat{\boldsymbol\sigma}_{12})\hat{\boldsymbol\sigma}_{12} - \Delta\hat{\bm\sigma}_{12}^\perp\,,\\
        \boldsymbol c'_2&=\boldsymbol C-\dfrac{1}{2}\boldsymbol c_{12}+\dfrac{1+\alpha}{2}(\boldsymbol c_{12}\cdot\hat{\boldsymbol\sigma}_{12})\hat{\boldsymbol\sigma}_{12}+\Delta\hat{\bm\sigma}_{12}^\perp\,.
    \end{split}
    \label{eq: change of variable}
\end{equation}
The integrals are Gaussian and evaluated with SymPy~\cite{sympy} in the SM, by fixing
\begin{equation}
    \bm c_{12}=\begin{pmatrix}
        s\\0
    \end{pmatrix}, \quad \hat{\bm{\sigma}}=\begin{pmatrix}
        \cos(\Psi)\\\sin(\Psi)
    \end{pmatrix},\quad\bm C = \begin{pmatrix}
        C_x\\
        C_y
    \end{pmatrix} \,.
    \label{eq: useful parametrization}
\end{equation}
As an example, we analytically calculate the integral $\omega\langle\delta E \rangle_{\rm coll}={\sigma\chi}{n^{-1}}\int \Theta(-\bm {v}_{12}\cdot \hat{ \bm\sigma}_{12})|\bm {v}_{12}\cdot \hat{ \bm\sigma}_{12}|\delta E f(\bm v_2)f(\bm v_1)d\hat{\bm\sigma}_{12}d\bm v_1 d \bm v_2$, given in Eq.~\eqref{eq: temperature}, which has the form of the previously discussed integral. We proceed by performing the change of variable Eqs.~\eqref{eq: change of variable}, which leads to
\begin{equation}
f(\bm v_1)f(\bm v_2)=\left(\frac{mn}{2\pi T}\right)^{2}
\exp\left(-\frac{m\bm C^{2}}{T}\right)\exp\left(-\frac{m\bm c_{12}^{2}}{4T}\right)\,.
\end{equation}
Then, using the parametrization Eq.~\eqref{eq: useful parametrization} yields:
\begin{equation}
    \begin{split}
    \omega\langle\delta E\rangle_{\rm coll}&=\dfrac{\sigma \chi}{n}\left(\dfrac{mn}{2\pi T}\right)^2\int\Theta\big(-\cos(\Psi)\big) \big|s\cos(\Psi)\big|s \\
    &\times \delta Ee^{-m\bm C^{2}/T}e^{-ms^{2}/4T} d\Psi d\bm C d s \,.
    \end{split}
\end{equation}
By direct computation, we obtain $\delta E 
= m\Delta^2- m\Delta s\sin(\Psi)-m(1-\alpha^2)s^2\cos(\Psi)^2/4$. 
The term in $\sin(\Psi)$ integrates to 0, while
\begin{equation}
\begin{split}
&\int_{\pi/2}^{3\pi/2}\cos\Psi  d\Psi=-2\,,\quad\int_0^\infty s^4 e^{-a s^2}ds=\frac{3\sqrt{\pi}}{8}a^{-5/2}\,,
\\
&\int_{\pi/2}^{3\pi/2}\cos^3\Psi d\Psi=-\frac{4}{3}\,,\quad\int_0^\infty s^2 e^{-a s^2}ds=\frac{\sqrt{\pi}}{4}a^{-3/2}\,,
\end{split}
\end{equation}
with $a=m/(4T)$.
Putting everything together yields
\begin{equation}
\omega \langle \delta E\rangle_{\rm coll}=2\sqrt{\pi}\chi\sigma n\sqrt{\frac{T}{m}}\Big[m\Delta^{2}-(1-\alpha^{2})T\Big]\,.
\end{equation}

\subsection{Viscosities}\label{sec: viscosity}
For the viscosities, we need to compute:
\begin{equation}
    \begin{split}
        \mathsf L_{\parallel\parallel}=\mathsf L_{\perp\perp}\equiv&\Big\langle \big(f^{(0)}\big)^{-1}D_{ij}, \mathcal L[D_{ij}] \Big\rangle\,,\\
        \mathsf L_{\parallel\perp}=-\mathsf L_{\perp\parallel}\equiv&\Big\langle \big(f^{(0)}\big)^{-1}D_{ij}, \mathcal L[D^\perp_{ij}] \Big\rangle\,.
    \end{split}
\end{equation}
Using Eq.~\eqref{eq: kinetic integral}, this gives
\begin{equation}
    \begin{split}
    \mathsf L_{\parallel\parallel}=&\chi\sigma\int  \Theta(-\boldsymbol c_{12}\cdot \hat{\boldsymbol \sigma})(-\boldsymbol c_{12}\cdot \hat{\boldsymbol \sigma})f^{(0)}(\boldsymbol c_1)f^{(0)}(\boldsymbol c_2)\\
    &\qquad\times D_{ij}(\boldsymbol c_2) \delta(D_{ij})d\boldsymbol c_1 d\boldsymbol c_2d\hat{\boldsymbol\sigma}\,,\\
    \mathsf L_{\parallel\perp}=&\chi\sigma\int  \Theta(-\boldsymbol c_{12}\cdot \hat{\boldsymbol \sigma})(-\boldsymbol c_{12}\cdot \hat{\boldsymbol \sigma})f^{(0)}(\boldsymbol c_1)f^{(0)}(\boldsymbol c_2)\\
    &\qquad\times D_{ij}^\perp(\boldsymbol c_2) \delta(D_{ij})d\boldsymbol c_1 d\boldsymbol c_2d\hat{\boldsymbol\sigma}\,.
\end{split}
\end{equation}
After some algebra and using Eq.~\eqref{eq: change}, we find:
\begin{equation}
    \begin{split}
    \dfrac{D_{ij}(\bm c_2)\delta(D_{ij})}{m^2}=&\;(\boldsymbol c_1'\cdot \boldsymbol c_2)^2 + (\boldsymbol c_2'\cdot \boldsymbol c_2)^2 - (\boldsymbol c_1\cdot \boldsymbol c_2)^2\\
    &- (\boldsymbol c_2\cdot \boldsymbol c_2)^2-\dfrac{1}{2}\boldsymbol c^2_2\delta(\boldsymbol  c^2)\,,\\
    \dfrac{D_{ij}^\perp(\bm c_2)\delta(D_{ij})}{m^2}=&\;(\boldsymbol c_2^\perp\cdot\boldsymbol c_1')(\boldsymbol c_2\cdot\boldsymbol c_1') +(\boldsymbol c_2^\perp\cdot\boldsymbol c_2')(\boldsymbol c_2\cdot\boldsymbol c_2')\,,\\
    \end{split}
\end{equation}
with $\bm c^\perp = \bm\varepsilon\cdot\bm c$. After inserting Eq.~\eqref{eq: change of variable} and carrying out the Gaussian integrations, we obtain (see SM)
\begin{equation}
    \begin{split}
    \mathsf{L}_{\parallel\parallel}&=- \chi\sigma n^2T^2 \sqrt{\dfrac{\pi T}{m}}\left(\dfrac{m\Delta^2}{T} + (1 + \alpha)(7 - 3\alpha)\right)\,,\\
    \mathsf{L}_{\parallel\perp}&=2\pi \chi\sigma n^{2}T^{2}\Delta(1-\alpha)\,.
    \end{split}
    \label{eq: appendix before replacing T for viscosity}
\end{equation}

From Eq.~\eqref{eq: expression for all viscosities}, we have:
\begin{equation}
    \begin{split}
    \eta_s&=-4n^2T^3\dfrac{\mathsf{L}_{\parallel\parallel}}{\mathsf{L}_{\parallel\parallel}^2+\mathsf{L}_{\parallel\perp}^2}\,, \\
    \eta_o&=4n^2T^3\dfrac{\mathsf{L}_{\parallel\perp}}{\mathsf{L}_{\parallel\parallel}^2+\mathsf{L}_{\parallel\perp}^2}\,.
    \end{split}
    \label{eq: viscosity appendix before changing}
\end{equation}

After replacing $\mathsf{L}_{\parallel\parallel}$ and $\mathsf{L}_{\parallel\perp}$ by their values and replacing $T$ by its steady-state value, we recover the main text result Eqs.~\eqref{eq: viscosity theory both}.

We also note that the odd-viscosity
\begin{equation}
    \eta_o = \dfrac{8\pi \chi \sigma n^4T^5}{\mathsf{L}_{\parallel\parallel}^2+\mathsf{L}_{\parallel\perp}^2}\Delta (1 - \alpha)\,,
    \label{eq: odd visco vanish before}
\end{equation}
vanishes in the limit $\alpha\to 1$, even \textit{before} setting $T$ to its steady-state value, as argued in the main text.

\subsection{Thermal conductivities}\label{sec: conductivity}

Similarly to the viscosity calculation, the thermal sector requires the matrix elements
\begin{equation}
    \mathsf M_{\alpha\beta}
    \equiv
    \Big\langle \big(f^{(0)}\big)^{-1}A^{(\alpha)}_{i}, \mathcal L[A^{(\beta)}_{i}] \Big\rangle\,,
    \quad
    \alpha,\beta\in\{\parallel,\perp\}\,,
\end{equation}
with $A_i^{(\parallel)}\equiv A_i$ and $A_i^{(\perp)}\equiv A_i^\perp$. Projecting the thermal part of Eq.~\eqref{eq: to solve} onto $\{ \mathbf A,\mathbf A^\perp\}$ gives

\begin{equation}
{\renewcommand{\arraystretch}{1.1}
\dfrac{1}{T}\begin{pmatrix}
   4nT^3/m\\
   0
\end{pmatrix}
=
\begin{pmatrix}
    \mathsf M_{\parallel\parallel} & \mathsf M_{\parallel\perp} \\
    \mathsf M_{\perp\parallel}     & \mathsf M_{\perp\perp}
\end{pmatrix}
\begin{pmatrix}
    \mathcal A_0 \\
    \mathcal A_0^\perp
\end{pmatrix}\,.
}
\end{equation}
where we used $\langle A_i,A_j\rangle=(2nT^3/m)\delta_{ij}$ and $\langle A_i^\perp,A_i\rangle=0$. The inversion yields:
\begin{equation}
    \mathcal A_0=\dfrac{4nT^2}{m}\dfrac{\mathsf{M}_{\parallel\parallel}}{\mathsf{M}_{\parallel\parallel}^2+\mathsf{M}_{\parallel\perp}^2}\,,\quad \mathcal A_0^\perp=\dfrac{4nT^2}{m}\dfrac{\mathsf{M}_{\parallel\perp}}{\mathsf{M}_{\parallel\parallel}^2+\mathsf{M}_{\parallel\perp}^2}\,.
    \label{eq: expression for A0}
\end{equation}

The matrix elements follow from the kinetic integral~\eqref{eq: kinetic integral} with integrands:
\begin{equation}
    \begin{split}
    \dfrac{A_{i}(\bm c_2)\delta(A_{i})}{m/2}&=\Big(\frac{m\boldsymbol c_2^2}{2}-2T\Big) \Big[\boldsymbol c_1'^2(\boldsymbol c_2\cdot\boldsymbol c_1') +\boldsymbol c_2'^2(\boldsymbol c_2\cdot\boldsymbol c_2')\\
    &\qquad\qquad\qquad-\boldsymbol c_1^2(\boldsymbol c_2\cdot\boldsymbol c_1) -\boldsymbol c_2^2(\bm c_2\cdot \bm c_2) \Big]\,, \\
    \dfrac{A_{i}^\perp(\bm c_2)\delta(A_{i})}{m/2}&=\Big(\frac{m\boldsymbol c_2^2}{2}-2T\Big) \Big[\boldsymbol c_1'^2(\boldsymbol c_2^\perp\cdot\boldsymbol c_1') +\boldsymbol c_2'^2(\boldsymbol c_2^\perp\cdot\boldsymbol c_2')\\
    &\qquad\qquad\qquad -\boldsymbol c_1^2(\boldsymbol c_2^\perp\cdot\boldsymbol c_1) -\boldsymbol c_2^2(\bm c_2^\perp\cdot \bm c_2) \Big]\,.
    \end{split}
\end{equation}
Upon integration, we obtain (see SM)
\begin{equation}
    \begin{split}
    \mathsf{M}_{\parallel\parallel}&=-\frac{\sqrt{\pi} T^{5/2} \chi n^{2} \sigma}{2 m^{3/2}}\Big(T(1 + \alpha)(19 - 15\alpha)-3m\Delta^{2}\Big)\,,\\
    \mathsf{M}_{\parallel\perp}&=\pi \chi \sigma n^2T^3 \Delta(1-\alpha)/m\,.
    \end{split}
\end{equation}

To relate $\mathsf M$ to the conductivities, we use the heat current 
\begin{equation}
    \begin{split}
    \bm J =& \dfrac{m}{2}\int \bm c^2\bm cf(\bm c)d\bm c\\
    =&\mu\dfrac{m}{2T}\int \bm c^2\bm cf^{(0)}(\bm c)\Big(\mathcal A_0\mathbf A +\mathcal A^\perp_0\mathbf A^\perp\Big)\cdot\bm \nabla Td\bm c\\
    =&\mu\left(\dfrac{2nT^2}{m}\mathcal A_0\bm 1-\dfrac{2nT^2}{m}\mathcal A_0^\perp\bm \varepsilon\right)\bm\nabla T\,,
    \end{split}
\end{equation}
so that
\begin{equation}
    \kappa=-\dfrac{2nT^2}{m}\mathcal A_0, \qquad \kappa_o=\dfrac{2nT^2}{m}\mathcal A_0^\perp\,.
\end{equation}
Substituting Eq.~\eqref{eq: expression for A0} and the steady-state temperature yields
\begin{flalign}
\label{eq:kappa_expr}
    &\kappa=\frac{16|\Delta|(1 + \alpha)(4-3\alpha)}{\sqrt{\pi}\chi \sigma \sqrt{1-\alpha^{2}}Q(\alpha)},\\
    &\kappa_o=\frac{8 \Delta \left(1-\alpha \right)}{\chi \sigma Q(\alpha)}\,,
\end{flalign}
with $Q(\alpha)$ a positive polynomial over $0\leq \alpha \leq 1$:
\begin{equation}
Q(\alpha)=\pi (1-\alpha)^{2} (1-\alpha^{2}) + 4 (1 + \alpha)^2(4-3\alpha)^{2}\,.
\end{equation}

\subsection{Self-diffusion coefficient}\label{sec: diffusivity}

We compute the regular self-diffusivity $D$ and its odd counterpart $D_o$ from the Boltzmann-Lorentz equation for a tracer following Ref.~\onlinecite{brilliantov2010kinetic}:
\begin{equation}
    \partial_t f_s(\mathbf r, \bm v, t)  + \bm v\cdot \bm \nabla f_s(\mathbf r, \bm v, t)= {\mathcal J}^{\rm dilute}\big(\mathbf r, \bm v|f_s, f_B\big)\,,
    \label{eq: Boltzmann Lorentz}
\end{equation}
where $f_B=f^{(0)}$ is a fixed \textit{homogeneous} Gaussian and only $f_s$ evolves with time. The tracer density $n_s(\mathbf r,t)=\int f_s(\mathbf r,\bm v,t)d\bm v$ obeys:
\begin{equation}
    \partial_tn_s = -\bm \nabla\cdot\bm J_s\,, \quad \bm J_s=-D\bm\nabla n_s-D_o(\bm\varepsilon\cdot\bm\nabla)n_s\,.
    \label{eq: diffusivity current appendix}
\end{equation}

We use a Chapman-Enskog expansion with, for convenience, a slightly modified ansatz relative to the main text
\begin{equation}
    f_s(\mathbf r, \bm v, t) = f_{s}^{(0)}[n_s(\mathbf r, t)|\bm v] + \mu f^{(1)}_s[n_s(\mathbf r, t)|\bm v].
\end{equation}
Since tracer and bath particles are identical, their homogeneous distribution must be the same:
\begin{equation}
    f_{s}^{(0)}=\dfrac{n_s(\mathbf r, t)}{n}f_B\,,
\end{equation}
where $n=(N-1)/L^{2}\simeq N/L^{2}$ is the global bath number density. At order $\mu^1$, the Boltzmann-Lorentz equation reads:
\begin{equation}
\partial_t^{(0)}f^{(1)}_s+\partial_t^{(1)}f_{s}^{(0)}+\bm v\cdot\bm\nabla f_{s}^{(0)}={\mathcal J}^{\rm dilute}(f_B, f^{(1)}_s)\,.
\end{equation}
In the steady state $\partial_t^{(0)}f^{(1)}_s=0$, and since $\partial_t n_s=\mathcal O(\bm\nabla^{2})$ we also have $\partial_t^{(1)}f_{s}^{(0)}=0$, yielding:
\begin{equation}
    \dfrac{1}{n}(\bm v\cdot\bm\nabla n_s)f_B={\mathcal J}^{\rm dilute}(f_B, f^{(1)}_s)\,.
\end{equation}

By linearity of ${\mathcal J}^{\rm dilute}(f_B, \cdots)$, we seek a solution of the form:
\begin{equation}
    f^{(1)}_s= (B(\bm v) \bm v\cdot\bm\nabla n_s+B^\perp(\bm v) \bm v^\perp\cdot\bm\nabla n_s)f_B(\bm v)\,,
\end{equation}
with $\bm v^\perp=\bm\varepsilon\cdot\bm v$. We assume again $B(\bm v)=B_0$ and $B^\perp(\bm v)=B_0^\perp$. Therefore, we must solve:
\begin{equation}
    \dfrac{1}{n}f_B\bm v=B_0{\mathcal J}^{\rm dilute}(f_B, f_B\bm v)+B_0^\perp {\mathcal J}^{\rm dilute}(f_B, f_B\bm v^\perp)\,.
\end{equation}
Projecting onto $\{\bm v,\bm v^\perp\}$ yields
\begin{equation}
    \dfrac{1}{n}
    \begin{pmatrix}
        2nT/m\\0
    \end{pmatrix}
    =
    \begin{pmatrix}
        \mathsf N_{\parallel\parallel}&\mathsf N_{\parallel\perp}\\
        -\mathsf N_{\parallel\perp}&\mathsf N_{\parallel\parallel}\\
    \end{pmatrix}
    \begin{pmatrix}
        B_0\\B_0^\perp
    \end{pmatrix},
    \label{eq: diffusivity Matrix}
\end{equation}
with (see SM)
\begin{equation}
    \begin{split}
    \mathsf N_{\parallel\parallel}=&\int \bm v\cdot {\mathcal J}^{\rm dilute}(f_B, f_B\bm v) d\bm v\\
    =&-2\sqrt{\pi (T/m)^3}\chi n^2\sigma (1 + \alpha)\,,\\
    \mathsf N_{\parallel\perp}=&\int \bm v\cdot {\mathcal J}^{\rm dilute}(f_B, f_B\bm v^\perp) d\bm v\\
    =&\dfrac{\pi\sigma \chi n^2T}{m}\Delta\,.
    \end{split}
\end{equation}

We now relate $B_0$ and $B^\perp_0$ to the diffusivities via $\bm J_s=\int \bm vf_sd\bm v$:
\begin{equation}
    \begin{split}
        \bm J_s =& \mu\int \bm vf_B(\bm v)\Big(B_0  \bm v + B_0^\perp \bm v^\perp\Big)\cdot\bm \nabla n_s d\bm v\\
        =&\mu\dfrac{nT}{m}\big(B_0\bm 1 - B_0^\perp\bm\varepsilon\big)\cdot\bm \nabla n_s\,.
    \end{split}
\end{equation}
By identification with Eq.~\eqref{eq: diffusivity current appendix}, we get:
\begin{equation}
    D = -\dfrac{nT}{m}B_0\,,\quad D_o=\dfrac{n T}{m}B_0^\perp\,.
\end{equation}
Inverting Eq.~\eqref{eq: diffusivity Matrix} and using the steady state temperature Eq.~\eqref{eq: temperature iso} yields:
\begin{equation}
    \begin{split}
        D=&\dfrac{\sqrt{\pi}\sigma|\Delta| }{\phi \chi\sqrt{1 - \alpha^2}R(\alpha)}\,,\\
        D_o=&\dfrac{\pi\sigma\Delta}{2\phi\chi (1 + \alpha) R(\alpha)}\,,
    \end{split}
\end{equation}
with $R(\alpha)$ a strictly positive polynomial over $0\leq \alpha \leq 1$:
\begin{equation}
    R(\alpha)=4(1 + \alpha) + \pi(1 - \alpha)\,.
\end{equation}

\section*{Supplementary Material}

A Python notebook is provided as Supplementary Material to automatically evaluate all integrals encountered in this work.

\bibliography{bib}

\end{document}